\documentclass[reprint,
groupedaddress,
superscriptaddress,
amsmath,amssymb,
aps,
]{revtex4-2}

\usepackage{graphicx}

\usepackage{dcolumn}

\usepackage{bm}

\usepackage{lineno}

\usepackage{graphicx}

\usepackage{dcolumn}

\usepackage{bm}

\usepackage{hyperref}

\usepackage{braket}

\usepackage{lipsum}

\usepackage{comment}

\usepackage{physics}

\usepackage[utf8]{inputenc}

\usepackage{xcolor}

\usepackage{empheq}

\usepackage{tikz}

\usepackage[normalem]{ulem}

\DeclareMathAlphabet\mbc{OMS}{cmsy}{b}{n}

\usepackage{balance}

\begin{document}

\global\long\def\eqn#1{\begin{align}#1\end{align}}

\global\long\def\vec#1{\overrightarrow{#1}}

\global\long\def\ket#1{\left|#1\right\rangle }

\global\long\def\bra#1{\left\langle #1\right|}

\global\long\def\bkt#1{\left(#1\right)}

\global\long\def\sbkt#1{\left[#1\right]}

\global\long\def\cbkt#1{\left\{#1\right\}}

\global\long\def\abs#1{\left\vert#1\right\vert}

\global\long\def\cev#1{\overleftarrow{#1}}

\global\long\def\der#1#2{\frac{{d}#1}{{d}#2}}

\global\long\def\pard#1#2{\frac{{\partial}#1}{{\partial}#2}}

\global\long\def\re{\mathrm{Re}}

\global\long\def\im{\mathrm{Im}}

\global\long\def\dd{\mathrm{d}}

\global\long\def\ddd{\mathcal{D}}

\global\long\def\avg#1{\left\langle #1 \right\rangle}

\global\long\def\mr#1{\mathrm{#1}}

\global\long\def\mb#1{{\mathbf #1}}

\global\long\def\mc#1{\mathcal{#1}}

\global\long\def\tr{\mathrm{Tr}}

\global\long\def\dbar#1{\Bar{\Bar{#1}}}

\global\long\def\nth{$n^{\mathrm{th}}$\,}

\global\long\def\mth{$m^{\mathrm{th}}$\,}

\global\long\def\non{\nonumber}

\newcommand{\orange}[1]{{\color{orange} {#1}}}

\newcommand{\cyan}[1]{{\color{cyan} {#1}}}

\newcommand{\yellow}[1]{{\color{yellow} {#1}}}

\newcommand{\green}[1]{{\color{green} {#1}}}

\newcommand{\red}[1]{{\color{red} {#1}}}

\global\long\def\todo#1{\orange{{$\bigstar$ \cyan{\bf\sc #1}}$\bigstar$} }

\preprint{APS/123-QED}

\title{Bottom-up Fabrication of 2D Rydberg Exciton Arrays in Cuprous Oxide}

\author{Kinjol Barua}
\email{kbarua@purdue.edu}

\affiliation{Elmore Family School of Electrical and Computer Engineering, Purdue University, West Lafayette, IN 47906, USA}

\author{Samuel Peana}

\affiliation{Elmore Family School of Electrical and Computer Engineering, Purdue University, West Lafayette, IN 47906, USA}

\author{Arya Deepak Keni}

\affiliation{Elmore Family School of Electrical and Computer Engineering, Purdue University, West Lafayette, IN 47906, USA}

\author{Vahagn Mkhitaryan}

\affiliation{Elmore Family School of Electrical and Computer Engineering, Purdue University, West Lafayette, IN 47906, USA}

\author{Vladimir Shalaev}

\affiliation{Elmore Family School of Electrical and Computer Engineering, Purdue University, West Lafayette, IN 47906, USA}

\author{Yong P. Chen}

\affiliation{Elmore Family School of Electrical and Computer Engineering, Purdue University, West Lafayette, IN 47906, USA}

\affiliation{Department of Physics and Astronomy, Purdue University, West Lafayette, IN 47906, USA}

\author{Alexandra Boltasseva}

\affiliation{Elmore Family School of Electrical and Computer Engineering, Purdue University, West Lafayette, IN 47906, USA}

\author{Hadiseh Alaeian}

\email{halaeian@purdue.edu}

\affiliation{Elmore Family School of Electrical and Computer Engineering, Purdue University, West Lafayette, IN 47906, USA}

\affiliation{Department of Physics and Astronomy, Purdue University, West Lafayette, IN 47906, USA}


\begin{abstract}
Solid-state platforms provide exceptional opportunities for advancing on-chip quantum technologies by enhancing interaction strengths through coupling, scalability, and robustness. Cuprous oxide (\( \rm Cu_{2}O \)) has recently emerged as a promising medium for scalable quantum technology due to its high-lying Rydberg excitonic states, akin to those in hydrogen atoms. To harness these nonlinearities for quantum applications, the confinement dimensions must match the Rydberg blockade size, which can reach several microns in \( \rm Cu_{2}O \).
Using a CMOS-compatible growth technique, this study demonstrates the bottom-up fabrication of site-selective arrays of \( \rm Cu_{2}O \) microparticles. We observed Rydberg excitons up to the principal quantum number \( n \)=5 within these \( \rm Cu_{2}O \) arrays on a quartz substrate and analyzed the spatial variation of their spectrum across the array, showing robustness and reproducibility on a large chip. These results lay the groundwork for the deterministic growth of \( \rm Cu_{2}O \) around photonic structures, enabling substantial light-matter interaction on integrated photonic platforms and paving the way for scalable, on-chip quantum devices.
\end{abstract}

\maketitle

%\tableofcontents

\section{Introduction}~\label{Introduction}

Achieving strong photon-photon interaction is a long-sought goal in quantum optics, holding the promise of transforming quantum information technologies. Photons emerge as compelling contenders for applications in quantum technology due to their resilience against environmental disruptions and capacity for long-distance transmission with minimal losses~\cite{Bouwmeester1997, O'Brien2007, Li2022, Maring2024}. Nevertheless, the absence of photon-photon interaction in linear optical media presents a formidable challenge in constructing extensive optical quantum communication networks and scalable quantum devices~\cite{O'Brien2007, Matthews2009, Aspuru-Guzik2012, Wang2020, Vigliar2021, Bourassa2021}. For example, strong photon-photon interactions are indispensable to perform two-qubit gate operations in photonic quantum circuits~\cite{O'Brien2003, Smith2009, Crespi2011, Barz2014, Shi2022} and to execute quantum algorithms on a photonic chip using entangled photon pairs~\cite{Politi2009}. Interactions amongst photons can be obtained in nonlinear media such as Beta-barium borate~\cite{Bache2013}, Lithium Niobate~\cite{Li1997}, and Potassium Niobate~\cite{Ludlow2001}. However, the $\chi^{(3)}$ nonlinear susceptibility of these materials is typically very small ($\sim$ $10^{-20}$ $m^2/V^2$) to obtain interactions at a single- or few-photon level. Therefore, an alternative scheme is necessary to achieve substantial photon-photon interactions.

Recently, Rydberg atoms and atom arrays have emerged as promising candidates for manifesting significant optical nonlinearity~\cite{Moreno2021, Mu2021, Chen2021, Bai:16} at a few photon levels. Rydberg atoms are highly excited electronic states with a large principal quantum number ($n$), and they offer exotic properties like longer coherence time~\cite{Hermann2014, Lampen2018, Contat2020} and extended wavefunction (scales as $n^2$) giving rise to very strong, long-range dipolar and van der Waals (vdW) interactions scaling as $n^4$ and $n^{11}$, respectively~\cite{Gallagher1994, Adams2020}. Besides, Rydberg states are highly tunable via external electric and magnetic fields~\cite{Jiao2022} finding various applications in quantum computing and information~\cite{Saffman2010}, quantum optics~\cite{Peyronel2012, Ripka2018, Kumlin_2023}, programmable quantum simulators~\cite{Bernien2017, Scholl2021, Bluvstein2022, Giudici2022, Nishad2023, Moss2024, Bluvstein2024, Michel2024}, precision classical and quantum sensing~\cite{Larrouy2019, Simons2021, Simons2021_2, Yuan2023} and metrology~\cite{Ovsiannikov2022}. While substantial advancements have been achieved in Rydberg atom quantum technologies~\cite{Fahey2011, Gorniaczyk2014, Ryabtsev2016, Saffman2016, Ripka2018, Levine2018, Duspayev2021}, a solid-state Rydberg platform with Rydberg excitons could be very advantageous due to its lower technical complexity and inherent integration capabilities. Solid-state systems provide a robust and miniaturized alternative, with sample sizes typically in the micron range, enabling them to fit into very compact experimental setups. This facilitates the easy tuning and control of individual excitons and allows for straightforward, controllable scalability. Additionally, with their inherent nature as an open quantum system, such platforms will make a unique testbed for the study of non-equilibrium quantum many-body physics~\cite{Delteil2019, Deligiannis2022, Bloch2022}.

\begin{figure*}[htpb!]
\centering
\includegraphics[width=18cm]{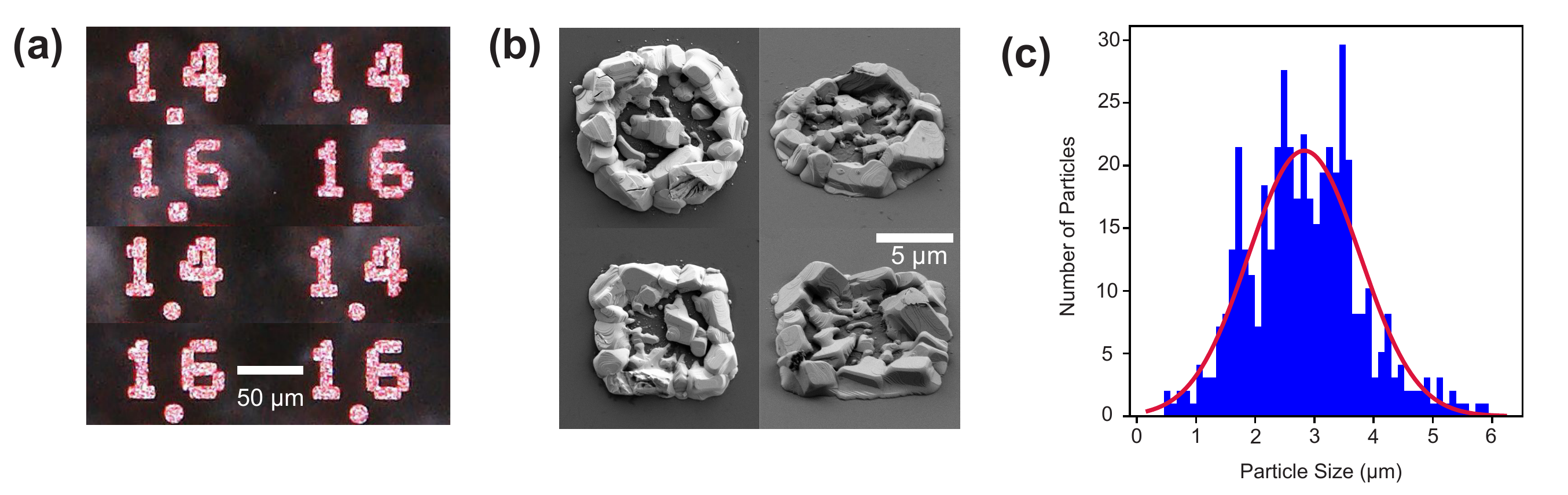}
\caption{
\textbf{Summary of optical and scanning electron microscope (SEM) characterization of the  Cu$\rm_2$O arrays.} (a) Optical microscope images of circular and square $\rm Cu_{2}O$ regions with nominal radii and side lengths of 14 and 16 $\mu$m, respectively. The $\rm Cu_{2}O$ regions showed Ruby color like natural gemstones. (b) SEM images of the same $\rm Cu_{2}O$ regions demonstrated the two-dimensional growth of crystalline, terrace-like micro-particles. (c) Size distribution of 500 $\rm Cu_{2}O$ micro-particles in various array sites showed a normal distribution with a mean size around 2.8 $\mu$m.}
\label{SampleImageFig}    
\end{figure*}

So far, Rydberg excitons have been observed in a few materials, such as two-dimensional monolayer transition metal dichalcogenides~\cite{Xiao2017, Mueller2018, Riis2020, Chaves2023} and perovskites~\cite{Bao2019, Ahumada-Lazo_2021, Su2021}. Among them, cuprous oxide ($\rm Cu_{2}O$) is the only semiconductor where Rydberg states up to $n$ = 30, natural bulk crystals, have been observed at cryogenic temperatures~\cite{Versteegh2021}. Cuprous oxide is a direct bandgap semiconductor ($E_g$=2.17 eV), and its electronic transitions from different valence bands to different conduction bands give rise to four distinct Rydberg exciton series (yellow, green, blue, and violet series) named after the color of photons they emit~\cite{Takahata2018}. In the yellow series, both the conduction band and valence band have the same parity, resulting in dipole-allowed p-excitons along with long-lived quadrupole-allowed yellow 1s-orthoexciton state~\cite{Jang2015, Asman2020}, making it significantly distinct from other excitonic systems. Furthermore, the large binding energy of $\rm Cu_{2}O$ ($\sim$ 98 meV) facilitates the observation of high-lying excitonic states without succumbing to thermal ionization~\cite{Kazimierczuk2014}.

Although natural bulk $\rm Cu_{2}O$ crystals can support Rydberg excitons, both in an unboundedcrystal~\cite{Takahata2018, KITAMURA2017808, Takahata2018May, NAKA2001413} and within confined pillars~\cite{Paul2024}, impurities and oxygen/copper vacancies negatively affect the quality of these excitons~\cite{Koirala2013}. Therefore, producing controlled, high-quality synthetic $\rm Cu_{2}O$ films is desirable. This approach also allows precise control over film thickness, making it comparable to the \emph{Rydberg blockade radius}, where only a single exciton is allowed, thus enabling extreme nonlinearity at the single-photon level.
Utilizing recently developed CMOS-compatible growth techniques~\cite{Steinhauer2020, DeLange2023}, we employed a thermal oxidation method for the bottom-up fabrication of 2D arrays of $\rm Cu_{2}O$ for the first time. Through non-resonant photoluminescence (PL) spectroscopy, we demonstrated the robustness of the fabricated arrays as well as the reproducibility and consistency of the Rydberg excitonic levels across $3000 \times 600$ $\mu m^2$-large chips. 
Combined with locally addressable techniques such as spatial confinement~\cite{Belov2024}, strain~\cite{Moon2020}, electric fields~\cite{Thureja2022}, and hybridizations with photons in structured microcavities~\cite{Bajoni2008, Kuznetsov2018, Orfanakis2022}, this bottom-up fabrication scheme establishes a foundation for realizing versatile programmable Rydberg exciton arrays, akin to Rydberg atom arrays, which pave the way for advancements in scalable, on-chip, and highly tunable quantum technologies.

\section{Results and Discussions}~\label{Results}
\begin{figure*}[htpb!]
\centering\includegraphics[width=18cm]{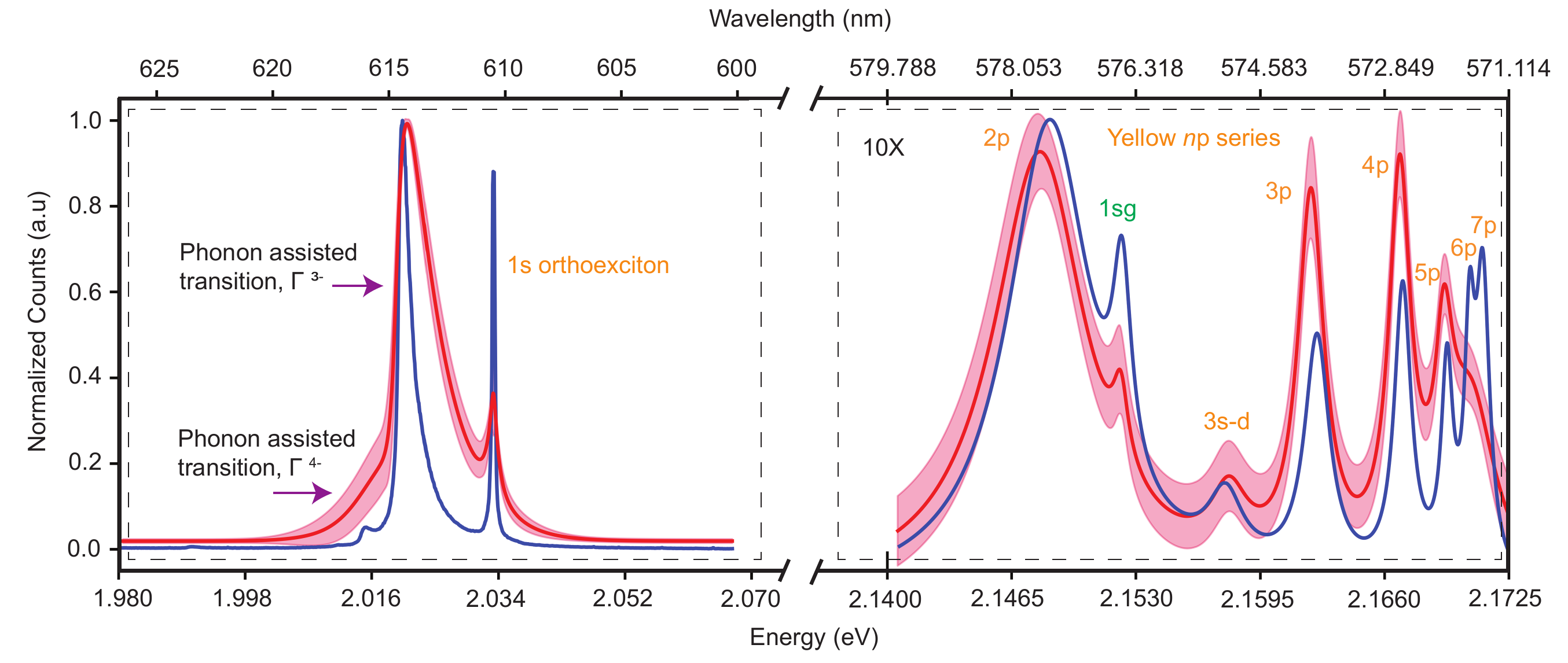}
\caption{\textbf{Summary of the results from low-temperature photoluminescence (PL) measurement from various $\rm Cu_{2}O$ regions (red curve) and thin-film $\rm Cu_{2}O$ sample (blue curve) having same thicknesses and grown under the same condition.} The left-hand side of the figure (600 - 625 nm) illustrates $\Gamma_3^-$ and $\Gamma_4^-$ phonon-assisted transitions, along with the yellow 1s-orthoexciton transition, whereas the right-hand side (570 - 580 nm) displays the photoluminescence (PL) from the yellow $n$p Rydberg exciton series. The $\Gamma_3^-$ phonon-assisted transition in the thin-film sample exhibits a narrower linewidth and a more pronounced 1s-orthoexciton transition compared to the arrays, where the 1s-orthoexciton transition is relatively broader and weaker in intensity. In the yellow exciton series, exciton peaks up to $n$=5p (7p) were observed for arrays (thin-film) in the PL spectrum. The 1s transition of the green series (denoted as 1sg) and 3s-d transitions are also discernible in the PL of the yellow exciton series. The yellow exciton spectrum is magnified by one order of magnitude to adequately display it alongside phonon replicas. The red-shaded region indicates the area of statistical uncertainty across all measurements.}
\label{PLFig}
\end{figure*}

The fabrication was started with a 700 nm-thick Cu film patterned using photolithography, followed by a liftoff process. The Cu regions were oxidized at high temperature and low pressure to form $\rm Cu_{2}O$. In Fig.~\ref{SampleImageFig}(a), microscope images depict circular and square $\rm Cu_{2}O$ regions with nominal side lengths and diameters of 14 $\mu$m and 16 $\mu$m. The actual diameters and side lengths of the $\rm Cu_{2}O$ array sites deviated from the nominal values of the Cu regions. The thermal oxidation of Cu resulted in a 30\% increase in size (cf. Fig. S1 and Fig. S2 of the supplementary information). Figure~\ref{SampleImageFig}(b) presents SEM images illustrating the 2D terrace-like growth of $\rm Cu_{2}O$ around the periphery of the circular and square regions. In some sites, isolated $\rm Cu_{2}O$ single-crystalline micro-particles were observed at the center, which significantly contributed to the PL signal (cf. Fig. S3 (a) of the Supplementary information). Furthermore, Fig.~\ref{SampleImageFig}(c) shows the size distribution of $\rm Cu_{2}O$ particles across all regions with varying dimensions, indicating an average size of $\sim$ 2.8 $\mu m$ with a standard deviation of $\sim$ 0.8 $\mu m$.

The low-temperature non-resonant PL spectrum from the yellow Rydberg exciton $np$ series (570 - 580 nm) and phonon-assisted transitions ($\Gamma_3^-$ and $\Gamma_4^-$) along with yellow 1s orthoexciton (600 - 625 nm) are shown in Fig.~\ref{PLFig} for Cu$_2$O arrays. The red curve represents the average PL spectrum of $\rm Cu_{2}O$ arrays across all regions with varying nominal side lengths, whereas the shaded region delimits the measurements' statistical distribution. We noted that the PL signals in the $\rm Cu_{2}O$ array were remarkably stable, exhibiting minimal spectral diffusion and blinking effects. Moreover, the sample showed no evidence of degradation after repeated temperature cycling or under high laser power excitation.

For comparison, the PL spectrum of a 700 nm Cu$_2$O film fabricated using the thermal oxidation method under the same conditions is also shown with a solid blue line. Due to the lower count of yellow excitons compared to phonon-assisted transitions and yellow 1s orthoexciton, these counts were multiplied by 10 to be visible on the same scale. In the site-selective $\rm Cu_{2}O$ arrays, exciton peaks up to $5p$ or $6p$ in the yellow series were visible for all particle sizes examined in the experiments. As the energies of higher-order excitons approach the bandgap energy of Cu$_2$O ($E_g$=2.17 eV), and the oscillator strengths diminish with increasing the principal quantum number ($n$), the spectral lines overlap, complicating the resolution of individual peaks. The large linewidths of the p-excitons arose from exciton-phonon scattering originating from lattice vibrations and imperfections~\cite{Toyozawa1962}. The heterogeneity of the sample led to variations in the PL spectra across the sample. Additionally, the positions and linewidths of exciton peaks exhibited slight variations between particles within the same array and among different arrays (see Fig. S4 of the Supplementary information). 

As can be seen, the exciton peaks ($2p - 5p$) of the array sample were redshifted compared to the thin-film sample. Besides, the $\Gamma_3^-$ phonon-assisted transition in the thin-film sample demonstrated a narrower linewidth compared to the array sample, yielding a lower effective exciton temperature. The energy shift was much larger for the $2p$ peak because of stronger screening of Coulomb interactions in these states and hence larger quantum defect $\delta(n,T)$~\cite{Asman2020}. In Fig.~\ref{PLFig}, the first additional peak between $2p$ and $3p$ exciton transitions was attributed to the 1s peak of the green exciton series~\cite{Schweiner2017, Takahata2018}.  Additionally, the even-parity $3s-d$ transition was also visible in the yellow exciton series spectrum. These peaks consistently appeared across all photoluminescence spectra.

\begin{figure*}[htpb!]
\centering\includegraphics[width=18cm]{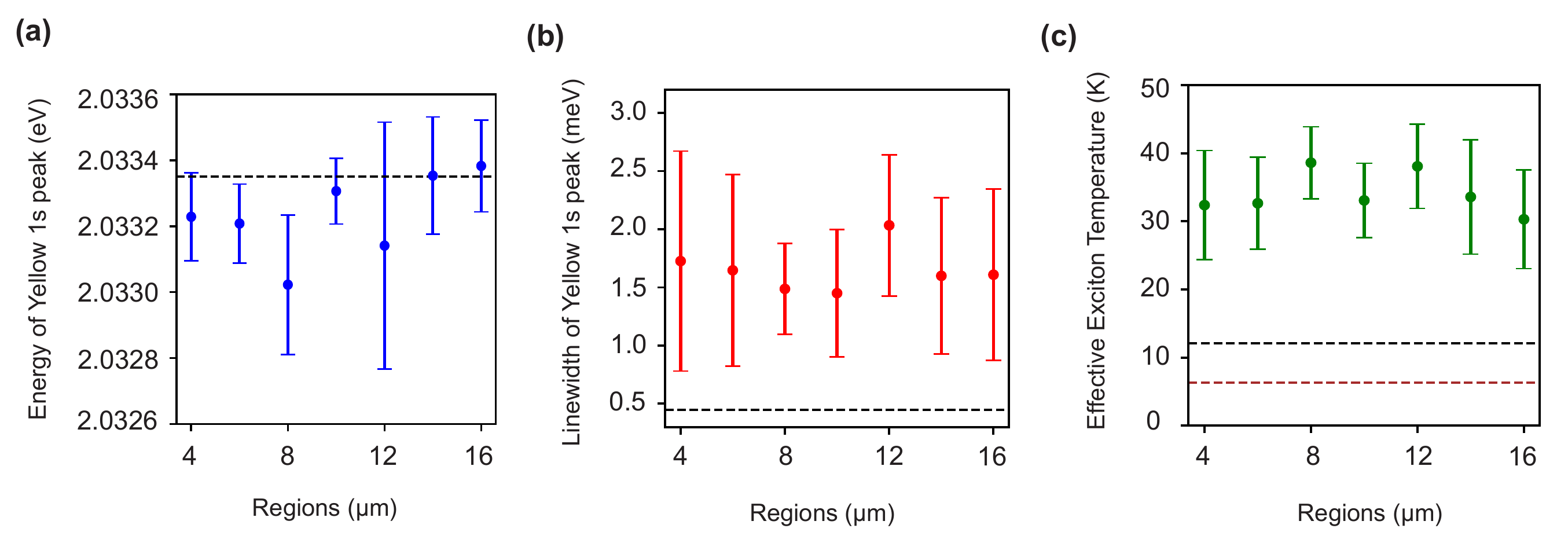}
\caption{\textbf{Statistical analysis of the exciton properties for $\rm Cu_{2}O$ micro-particle arrays.} (a) Energy of 1s-orthoexciton (statistical average over many arrays having the same size) from regions of all sizes. Here, for all regions, the energies of yellow 1s-orthoexciton peaks in arrays showed a redshift compared to the thin-film sample (black dashed line). (b) Linewidth (FWHM) of yellow 1s-orthoexciton from all $\rm Cu_{2}O$ arrays. The 1s-orthoexciton peak from the thin-film sample (black dashed line) demonstrated a narrower linewidth than arrays.  (c) The effective temperature of excitons from all regions extracted by fitting the phonon-assisted transition using a Maxwell-Boltzmann distribution function convoluted with a Gaussian. The black dashed line depicts the effective exciton temperature for the thin-film sample, whereas the brown dashed line is the temperature of the cryostat sample stage. The effective temperature of the excitons was different from the sample stage temperature, which was fixed at 6.3 K. On the other hand, for thin-film samples, the effective exciton temperature was much smaller, around 12.07 K. The sample temperature was different from the cryo stage temperature because of the poor thermal conductivity of the quartz substrate. The higher effective exciton temperatures for arrays contributed to the linewidth broadening and redshift of the 1s-orthoexciton peak. Error bars are statistical error bars for all figures.}
\label{StatFig}
\end{figure*}

For a quantitative study of the Rydberg excitons in the array, we used an asymmetric Fano lineshape to fit $\alpha_n(E)$ associated with the $n^{\textrm{th}}$ exciton transition~\cite{Kazimierczuk2014, Asman2020},

\begin{equation}
    \alpha_n(E) = f_n \frac{\frac{\Gamma_n}{2} + 2 q_n (E - E_n)}{\big(\frac{\Gamma_n}{2}\big)^2 + (E - E_n)^2}\, ,
\end{equation}

where $E_n$ is the energy of the $n^{\textrm{th}}$ transition, $\Gamma_n$ is the corresponding linewidth, $f_n$ is a scaling factor that is proportional to the oscillator strength, and $q_n$ is an asymmetry factor modeling the interference between the phonon continuum and narrow optical transitions~\cite{Toyozawa1964}. Utilizing this lineshape, we applied a least squares algorithm to fit the PL spectrum. 

In Fig.~\ref{PLFig}, in addition to the Rydberg series, the PL also encompassed a distinctly sharp yellow 1s-orthoexciton peak (approximately at 610 nm) alongside the $\Gamma_3^-$ and $\Gamma_4^-$ phonon-assisted relaxations of the 1s state characterized by a Boltzmann-tailed profile. We conducted a fitting procedure for the yellow 1s-orthoexciton peak, from which we derived both its energy and linewidth. The Boltzmann-like phonon-assisted transitions ($\Gamma_3^-$ and $\Gamma_4^-$) were convoluted with a Gaussian function to account for the spectrometer's instrumental response. Subsequently, the effective temperature ($T_{\rm eff}$) of the excitons was derived from the resultant fit parameters (cf. Supplementary section 2 and Fig. S3 (b)). The data depicted in Figure ~\ref{PLFig} reveals that a broader linewidth of the phonon-assisted transition ($\Gamma_3^-$) in arrays corresponded to a higher effective exciton temperature. Consequently, the phonon-assisted transition predominated over the yellow 1s orthoexciton, leading to decreased counts in the 1s state. This observation was drawn by comparing the red and blue curves, which corresponded to the array and thin-film samples, respectively. Furthermore, in the thin-film sample, the resolution extended up to $7p$ peaks within the yellow exciton series.

Figure~\ref{StatFig}(a) and (b) show the variations in both energy and linewidth of the yellow 1s-orthoexciton transition across the array sample, respectively. Each data point represents the averaged exciton energy obtained from a fitting procedure conducted on a set of 12 similar particles. The associated error bars signify the standard deviation across this ensemble. The energy and linewidth of the 1s-orthoexciton in a thin-film sample were illustrated by a horizontal black dashed line in both plots.

\begin{figure*}[htpb!]
\centering\includegraphics[width=18cm]{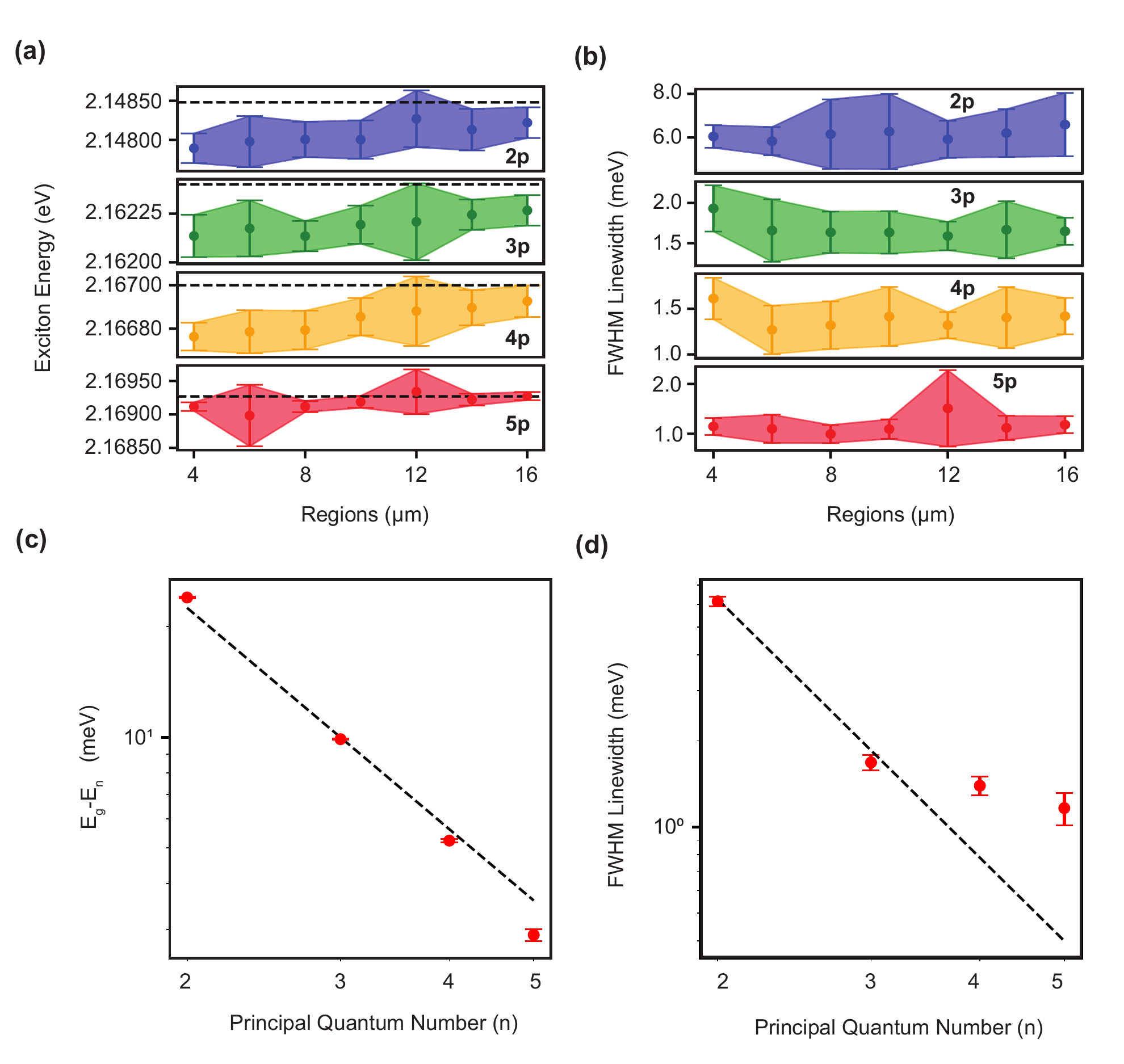}
\caption{\textbf{Statistical analysis of yellow Rydberg exciton energies for all $\rm Cu_{2}O$ arrays having nominal side lengths from 4 $\mu$m to 16 $\mu$m.}  (a)The variation in the energies of 2p, 3p, 4p, and 5p excitons was examined across arrays of different dimensions. The dashed line represents the exciton energies observed in a thin-film sample. In all instances, the exciton energies within the arrays exhibited a red shift relative to the thin-film sample. (b) FWHM linewidths of the 2p, 3p, 4p, and 5p excitons and their variation for all arrays of different sizes. (c) Rydberg exciton energies as a function of $n$ averaged over all regions with different nominal sidelengths from 4 $\mu$m to 16 $\mu$m. The black dashed line shows the $n^{-2}$ trend of the hydrogenic energy levels from bulk natural $\rm Cu_{2}O$ sample. The 2p, 3p, and 4p exciton energy followed the trendline, whereas the 5p exciton significantly deviated from it. (d) FWHM of yellow exciton series a function of principal quantum number (n). The black dashed line shows the $n^{-3}$ trend. For 4p and 5p exciton, their linewidths significantly deviated from the trendline and reached a plateau for higher principal quantum numbers. }
\label{FinalFig}
\end{figure*}

While Fig.~\ref{StatFig}(a) illustrates a redshift in the 1s-orthoexciton energies in $\rm Cu_{2}O$ arrays compared to the thin film, panel (b) shows that the linewidths of excitons in arrays are broader than those in the thin-film sample. Finally, in Fig.~\ref{StatFig}(c), the effective exciton temperature of the array sample was plotted alongside that of the thin-film sample (black dashed line) and the cryostation cold finger temperature (brown dashed line). For the thin-film sample, the effective exciton temperature of $\sim$ 12 K was closer to the platform, i.e., 6.3 K, with slight deviations attributed to the limited thermal conductivity between the cold finger and the sample due to the quartz substrate's poor thermal conductivity. Across all array sites, however, the effective exciton temperatures exceed the cryostation's platform temperature by 20 K to 40 K. Besides the aforementioned reasons for the discrepancy between the sample's and the cold finger's temperature, the higher exciton temperature in the array sample compared to the thin film can be attributed to the reduced thermal contact of the isolated Cu$_2$O islands with the substrate, leading to insufficient cooling. 
This local heating led to a spectral redshift of the resonant energies and the linewidth broadening in arrays compared to thin-film samples, as depicted in Fig.~\ref{StatFig}(a) and (b), respectively. It is worth mentioning that the modification of the energies, linewidth, and effective temperature revealed the uniformity among lattice sites despite the large sample size, confirming the robustness and overall consistency of the exciton properties.

To investigate the spatial heterogeneity of the Rydberg exciton properties across the sample, we presented the energies and full width at half maximum (FWHM) linewidths of the yellow exciton series of the array sample in Fig.~\ref{FinalFig}(a) and (b), respectively. The thin-film data are represented by the black dashed line in each panel. While the energies and linewidths of Rydberg excitons showed a generally similar trend across different sites and arrays, the remaining discrepancies were attributed to variations in effective exciton temperature across the sample. The enhanced uncertainties in the linewidth of the $5p$ state primarily resulted from fitting inaccuracies. Given the proximity of the $5p$ state energy to the bandgap energy of Cu$_2$O ($E_g=2.17$ eV) and potential overlap with other peaks, accurately determining the linewidth was challenging. Similar to the energy of the yellow 1s orthoexciton, the energies of the $np$ excitons in the Cu$_2$O array showed a systematic spectral redshift compared to the thin film due to local heating.

Finally, the dependence of Rydberg exciton energy and linewidths on the principal quantum number ($n$) is illustrated in Fig.~\ref{FinalFig}(c) and (d), respectively, obtained from the averaged data across all Cu$_2$O arrays. In Fig.~\ref{FinalFig}(c), the dashed black line shows the anticipated hydrogen-like energy scaling of $n^{-2}$ with the principal quantum number $n$. As can be seen, the observed exciton energies exhibited slight deviations from this power-law scaling. The notable difference in the $2p$ energy primarily arose from quantum defects attributable to the intensified Coulomb interaction with valence band holes. The deviations for higher states, e.g., $5p$, on the other hand, can arise due to the higher susceptibility of these states to the size and the non-spherical symmetry of the electron-hole potential. As a result, the energy of excitons exhibited deviations from the ideal hydrogenic series. In Fig.~\ref{FinalFig}(d), the linewidths of yellow excitons were plotted vs. the principal quantum number $n$. The linewidths of the excitons can be modeled as~\cite{Kazimierczuk2014}

\begin{equation}
    \Gamma_n= \alpha\frac{n^2 - 1}{n^5} + \beta\, ,
\label{LinewidthModel}
\end{equation}

where $\Gamma_n$ is the FWHM linewidth of the $n$th state and $\alpha$ is a parameter capturing the power-law scaling of the $p$-Rydberg states. For large $n$, this asymptotically approached the well-known scaling of $n^{-3}$ for $s$-type hydrogenic Rydberg states (black dashed trendline in panel (d)). On the other hand, $\beta$ is an empirical fitting parameter that models the observed plateau in several experiments~\cite{Kruger, Kazimierczuk2014, Schone2016}. As can be seen, the experimental linewidths reached a plateau following previous results~\cite{Kazimierczuk2014}. Although several hypotheses have been proposed for this behavior, including collisions with free electron plasma, dense clouds of ground-state excitons, or phonons, delineating the root cause requires a separate study in a sample where more Rydberg lines are attainable.

\section{Conclusion}~\label{conclusion}

In this study, we have demonstrated the first CMOS-compatible bottom-up fabrication of 2D cuprous oxide arrays on a quartz substrate and observed Rydberg exciton states up to $5p$ through non-resonant photoluminescence spectroscopy at cryogenic temperatures. Our robust fabrication method provides the monolithic site-selective growth of Cu$_2$O islands and lattice sites without noticeable degradation of the optical properties over time and after several cooling-heating cycles. We studied the spatial variations of the photoluminescence spectrum of Rydberg excitons and ensured the reproducibility of the results and robustness of the excitonic features across a large area of $3000 \times 600$ $\mu m^2$. 

This study opens up new avenues for creating arbitrary arrays of Rydberg excitons akin to Rydberg atom arrays. When combined with several established local addressability and fine-tuning schemes in solid-state platforms—such as laser power, AC/DC Stark shift, and surface acoustic waves—this platform offers a unique opportunity to study the non-equilibrium dynamics of lattice models with strong long-range interactions mediated by Rydberg excitons. The in-situ controllability of the lattice sites enables the dynamic mapping of optimization problems on an on-chip, programmable platform with favorable SWaP (size, weight, and power) features~\cite{Taylor_2022}. 
Moreover, the developed site-controlled grown Cu$_2$O can be integrated with nanophotonic circuitry and plasmonic nanostructures to mediate strong interactions between photons via Rydberg excitons and control the optical properties of these unique states using the advanced toolbox of nanophotonics~\cite{Neubauer2022, Ziemkiewicz2022}.

\section{Methods}~\label{Methods}
\subsection{Sample Preparation}~\label{sample_prep}
A 10 mm $\times$ 10 mm chip was cut from a 4-inch quartz wafer and cleaned via Toluene, Acetone, and IPA (Isopropyl Alcohol) baths, each lasting 5 minutes in a sonicator. Subsequently, the chip underwent drying using a nitrogen gun to eliminate any remaining solvents. A consistent layer of photoresist (AZ1518), approximately 4 $\mu$m in thickness, was applied to the chip via spin-coating, followed by a baking procedure. A photolithography design incorporating circular and square shapes, ranging in nominal size from 2 $\mu$m to 16 $\mu$m for each shape, was implemented. The photoresist was developed using the MF-26 photoresist developer for 60 seconds, followed by a rinse with distilled (DI) water.

Following the photolithography process, a 5 nm titanium (Ti) adhesion layer was deposited via an e-beam evaporation, which was succeeded by 700 nm of copper (Cu). After deposition, the photoresist layer was removed. Examination via scanning electron microscopy (SEM) of the copper sample revealed the presence of peripheral "wings" on both circular and square copper islands (Supplementary Fig. S1(a)), attributed to excessive copper accumulation on the angled sidewalls of the photoresist after liftoff. These wings were observed to diminish upon oxidation of the copper islands at elevated temperatures, forming Cu$_2$O.
The chip underwent oxidation within a CVD furnace for an extended duration to ensure thorough oxidation, as referenced in previous studies~\cite{Steinhauer2020, DeLange2023}. Following oxidation, all copper islands turned into cuprous oxide, evident from optical microscopy images and scanning electron microscope (SEM) images as illustrated in Fig.~\ref{SampleImageFig}(a) and (b), respectively.

\subsection{Photolumiscence Measurement}
A 532 nm CW laser was employed to assess the photoluminescence of the Cu$_2$O arrays at 6.3 K. This methodology facilitated an examination of the characteristics of both the yellow 1s orthoexcitons and the yellow $np$ excitons via PL investigations. The measurements were conducted under a constant power of 50 $\mu$W with a laser spot size around 3.3 $\mu$m (FWHM), with power stability maintained within $0.5\%$  using an acoustic optic modulator (AOM) configured in a double pass setup, as outlined in our previous work~\cite{DeLange2023}.

\begin{acknowledgments}
KB and HA acknowledge the financial support from the Industry-University Cooperative Research Center Program at the US National Science Foundation under grant No. 2224960. ADK and HA acknowledge the support from the Air Force Office of Scientific Research under award number FA9550-23-1-0489.
\end{acknowledgments}

\section*{Author contributions statement}

KB grew the Cu$_2$O sample, conducted the experiment, analyzed the data, and wrote the first draft of the manuscript. SP performed photolithography and liftoffs. KB and ADK performed the measurements and analyzed the data. VS, AB, YPC, and HA supervised the project. All authors contributed to writing and editing the manuscript.

\section*{Disclosures}
The authors declare no conflicts of interest.

\section*{Data Availability Statement}
The data presented in the paper is available at \url{https://zenodo.org/records/11608450}

\section*{Code Availability Statement}
The code used in the data analysis presented here is available from the authors upon reasonable request.

\newpage

\bibliography{apssamp}% Produces the bibliography via BibTeX.

\begin{thebibliography}{87}%
\makeatletter
\providecommand \@ifxundefined [1]{%
 \@ifx{#1\undefined}
}%
\providecommand \@ifnum [1]{%
 \ifnum #1\expandafter \@firstoftwo
 \else \expandafter \@secondoftwo
 \fi
}%
\providecommand \@ifx [1]{%
 \ifx #1\expandafter \@firstoftwo
 \else \expandafter \@secondoftwo
 \fi
}%
\providecommand \natexlab [1]{#1}%
\providecommand \enquote  [1]{``#1''}%
\providecommand \bibnamefont  [1]{#1}%
\providecommand \bibfnamefont [1]{#1}%
\providecommand \citenamefont [1]{#1}%
\providecommand \href@noop [0]{\@secondoftwo}%
\providecommand \href [0]{\begingroup \@sanitize@url \@href}%
\providecommand \@href[1]{\@@startlink{#1}\@@href}%
\providecommand \@@href[1]{\endgroup#1\@@endlink}%
\providecommand \@sanitize@url [0]{\catcode `\\12\catcode `\$12\catcode `\&12\catcode `\#12\catcode `\^12\catcode `\_12\catcode `\%12\relax}%
\providecommand \@@startlink[1]{}%
\providecommand \@@endlink[0]{}%
\providecommand \url  [0]{\begingroup\@sanitize@url \@url }%
\providecommand \@url [1]{\endgroup\@href {#1}{\urlprefix }}%
\providecommand \urlprefix  [0]{URL }%
\providecommand \Eprint [0]{\href }%
\providecommand \doibase [0]{https://doi.org/}%
\providecommand \selectlanguage [0]{\@gobble}%
\providecommand \bibinfo  [0]{\@secondoftwo}%
\providecommand \bibfield  [0]{\@secondoftwo}%
\providecommand \translation [1]{[#1]}%
\providecommand \BibitemOpen [0]{}%
\providecommand \bibitemStop [0]{}%
\providecommand \bibitemNoStop [0]{.\EOS\space}%
\providecommand \EOS [0]{\spacefactor3000\relax}%
\providecommand \BibitemShut  [1]{\csname bibitem#1\endcsname}%
\let\auto@bib@innerbib\@empty
%</preamble>
\bibitem [{\citenamefont {Bouwmeester}\ \emph {et~al.}(1997)\citenamefont {Bouwmeester}, \citenamefont {Pan}, \citenamefont {Mattle}, \citenamefont {Eibl}, \citenamefont {Weinfurter},\ and\ \citenamefont {Zeilinger}}]{Bouwmeester1997}%
  \BibitemOpen
  \bibfield  {author} {\bibinfo {author} {\bibfnamefont {D.}~\bibnamefont {Bouwmeester}}, \bibinfo {author} {\bibfnamefont {J.-W.}\ \bibnamefont {Pan}}, \bibinfo {author} {\bibfnamefont {K.}~\bibnamefont {Mattle}}, \bibinfo {author} {\bibfnamefont {M.}~\bibnamefont {Eibl}}, \bibinfo {author} {\bibfnamefont {H.}~\bibnamefont {Weinfurter}},\ and\ \bibinfo {author} {\bibfnamefont {A.}~\bibnamefont {Zeilinger}},\ }\bibfield  {title} {\bibinfo {title} {Experimental quantum teleportation},\ }\href {https://doi.org/10.1038/37539} {\bibfield  {journal} {\bibinfo  {journal} {Nature}\ }\textbf {\bibinfo {volume} {390}},\ \bibinfo {pages} {575} (\bibinfo {year} {1997})}\BibitemShut {NoStop}%
\bibitem [{\citenamefont {O'Brien}(2007)}]{O'Brien2007}%
  \BibitemOpen
  \bibfield  {author} {\bibinfo {author} {\bibfnamefont {J.~L.}\ \bibnamefont {O'Brien}},\ }\bibfield  {title} {\bibinfo {title} {Optical quantum computing},\ }\href {https://doi.org/10.1126/science.1142892} {\bibfield  {journal} {\bibinfo  {journal} {Science}\ }\textbf {\bibinfo {volume} {318}},\ \bibinfo {pages} {1567} (\bibinfo {year} {2007})}\BibitemShut {NoStop}%
\bibitem [{\citenamefont {Li}\ \emph {et~al.}(2022)\citenamefont {Li}, \citenamefont {Cao}, \citenamefont {Li}, \citenamefont {Cai}, \citenamefont {Liu}, \citenamefont {Ren}, \citenamefont {Liao}, \citenamefont {Wu}, \citenamefont {Li}, \citenamefont {Li}, \citenamefont {Liu}, \citenamefont {Lu}, \citenamefont {Yin}, \citenamefont {Chen}, \citenamefont {Peng},\ and\ \citenamefont {Pan}}]{Li2022}%
  \BibitemOpen
  \bibfield  {author} {\bibinfo {author} {\bibfnamefont {B.}~\bibnamefont {Li}}, \bibinfo {author} {\bibfnamefont {Y.}~\bibnamefont {Cao}}, \bibinfo {author} {\bibfnamefont {Y.-H.}\ \bibnamefont {Li}}, \bibinfo {author} {\bibfnamefont {W.-Q.}\ \bibnamefont {Cai}}, \bibinfo {author} {\bibfnamefont {W.-Y.}\ \bibnamefont {Liu}}, \bibinfo {author} {\bibfnamefont {J.-G.}\ \bibnamefont {Ren}}, \bibinfo {author} {\bibfnamefont {S.-K.}\ \bibnamefont {Liao}}, \bibinfo {author} {\bibfnamefont {H.-N.}\ \bibnamefont {Wu}}, \bibinfo {author} {\bibfnamefont {S.-L.}\ \bibnamefont {Li}}, \bibinfo {author} {\bibfnamefont {L.}~\bibnamefont {Li}}, \bibinfo {author} {\bibfnamefont {N.-L.}\ \bibnamefont {Liu}}, \bibinfo {author} {\bibfnamefont {C.-Y.}\ \bibnamefont {Lu}}, \bibinfo {author} {\bibfnamefont {J.}~\bibnamefont {Yin}}, \bibinfo {author} {\bibfnamefont {Y.-A.}\ \bibnamefont {Chen}}, \bibinfo {author} {\bibfnamefont {C.-Z.}\ \bibnamefont {Peng}},\ and\ \bibinfo {author} {\bibfnamefont {J.-W.}\ \bibnamefont {Pan}},\
  }\bibfield  {title} {\bibinfo {title} {Quantum state transfer over 1200 km assisted by prior distributed entanglement},\ }\href {https://doi.org/10.1103/PhysRevLett.128.170501} {\bibfield  {journal} {\bibinfo  {journal} {Phys. Rev. Lett.}\ }\textbf {\bibinfo {volume} {128}},\ \bibinfo {pages} {170501} (\bibinfo {year} {2022})}\BibitemShut {NoStop}%
\bibitem [{\citenamefont {Maring}\ \emph {et~al.}(2024)\citenamefont {Maring}, \citenamefont {Fyrillas}, \citenamefont {Pont}, \citenamefont {Ivanov}, \citenamefont {Stepanov}, \citenamefont {Margaria}, \citenamefont {Hease}, \citenamefont {Pishchagin}, \citenamefont {Lema{\^i}tre}, \citenamefont {Sagnes}, \citenamefont {Au}, \citenamefont {Boissier}, \citenamefont {Bertasi}, \citenamefont {Baert}, \citenamefont {Valdivia}, \citenamefont {Billard}, \citenamefont {Acar}, \citenamefont {Brieussel}, \citenamefont {Mezher}, \citenamefont {Wein}, \citenamefont {Salavrakos}, \citenamefont {Sinnott}, \citenamefont {Fioretto}, \citenamefont {Emeriau}, \citenamefont {Belabas}, \citenamefont {Mansfield}, \citenamefont {Senellart}, \citenamefont {Senellart},\ and\ \citenamefont {Somaschi}}]{Maring2024}%
  \BibitemOpen
  \bibfield  {author} {\bibinfo {author} {\bibfnamefont {N.}~\bibnamefont {Maring}}, \bibinfo {author} {\bibfnamefont {A.}~\bibnamefont {Fyrillas}}, \bibinfo {author} {\bibfnamefont {M.}~\bibnamefont {Pont}}, \bibinfo {author} {\bibfnamefont {E.}~\bibnamefont {Ivanov}}, \bibinfo {author} {\bibfnamefont {P.}~\bibnamefont {Stepanov}}, \bibinfo {author} {\bibfnamefont {N.}~\bibnamefont {Margaria}}, \bibinfo {author} {\bibfnamefont {W.}~\bibnamefont {Hease}}, \bibinfo {author} {\bibfnamefont {A.}~\bibnamefont {Pishchagin}}, \bibinfo {author} {\bibfnamefont {A.}~\bibnamefont {Lema{\^i}tre}}, \bibinfo {author} {\bibfnamefont {I.}~\bibnamefont {Sagnes}}, \bibinfo {author} {\bibfnamefont {T.~H.}\ \bibnamefont {Au}}, \bibinfo {author} {\bibfnamefont {S.}~\bibnamefont {Boissier}}, \bibinfo {author} {\bibfnamefont {E.}~\bibnamefont {Bertasi}}, \bibinfo {author} {\bibfnamefont {A.}~\bibnamefont {Baert}}, \bibinfo {author} {\bibfnamefont {M.}~\bibnamefont {Valdivia}}, \bibinfo {author} {\bibfnamefont {M.}~\bibnamefont
  {Billard}}, \bibinfo {author} {\bibfnamefont {O.}~\bibnamefont {Acar}}, \bibinfo {author} {\bibfnamefont {A.}~\bibnamefont {Brieussel}}, \bibinfo {author} {\bibfnamefont {R.}~\bibnamefont {Mezher}}, \bibinfo {author} {\bibfnamefont {S.~C.}\ \bibnamefont {Wein}}, \bibinfo {author} {\bibfnamefont {A.}~\bibnamefont {Salavrakos}}, \bibinfo {author} {\bibfnamefont {P.}~\bibnamefont {Sinnott}}, \bibinfo {author} {\bibfnamefont {D.~A.}\ \bibnamefont {Fioretto}}, \bibinfo {author} {\bibfnamefont {P.-E.}\ \bibnamefont {Emeriau}}, \bibinfo {author} {\bibfnamefont {N.}~\bibnamefont {Belabas}}, \bibinfo {author} {\bibfnamefont {S.}~\bibnamefont {Mansfield}}, \bibinfo {author} {\bibfnamefont {P.}~\bibnamefont {Senellart}}, \bibinfo {author} {\bibfnamefont {J.}~\bibnamefont {Senellart}},\ and\ \bibinfo {author} {\bibfnamefont {N.}~\bibnamefont {Somaschi}},\ }\bibfield  {title} {\bibinfo {title} {A versatile single-photon-based quantum computing platform},\ }\bibfield  {journal} {\bibinfo  {journal} {Nature Photonics}\
  }\href {https://doi.org/10.1038/s41566-024-01403-4} {10.1038/s41566-024-01403-4} (\bibinfo {year} {2024})\BibitemShut {NoStop}%
\bibitem [{\citenamefont {Matthews}\ \emph {et~al.}(2009)\citenamefont {Matthews}, \citenamefont {Politi}, \citenamefont {Stefanov},\ and\ \citenamefont {O'Brien}}]{Matthews2009}%
  \BibitemOpen
  \bibfield  {author} {\bibinfo {author} {\bibfnamefont {J.~C.~F.}\ \bibnamefont {Matthews}}, \bibinfo {author} {\bibfnamefont {A.}~\bibnamefont {Politi}}, \bibinfo {author} {\bibfnamefont {A.}~\bibnamefont {Stefanov}},\ and\ \bibinfo {author} {\bibfnamefont {J.~L.}\ \bibnamefont {O'Brien}},\ }\bibfield  {title} {\bibinfo {title} {Manipulation of multiphoton entanglement in waveguide quantum circuits},\ }\href {https://doi.org/10.1038/nphoton.2009.93} {\bibfield  {journal} {\bibinfo  {journal} {Nature Photonics}\ }\textbf {\bibinfo {volume} {3}},\ \bibinfo {pages} {346} (\bibinfo {year} {2009})}\BibitemShut {NoStop}%
\bibitem [{\citenamefont {Aspuru-Guzik}\ and\ \citenamefont {Walther}(2012)}]{Aspuru-Guzik2012}%
  \BibitemOpen
  \bibfield  {author} {\bibinfo {author} {\bibfnamefont {A.}~\bibnamefont {Aspuru-Guzik}}\ and\ \bibinfo {author} {\bibfnamefont {P.}~\bibnamefont {Walther}},\ }\bibfield  {title} {\bibinfo {title} {Photonic quantum simulators},\ }\href {https://doi.org/10.1038/nphys2253} {\bibfield  {journal} {\bibinfo  {journal} {Nature Physics}\ }\textbf {\bibinfo {volume} {8}},\ \bibinfo {pages} {285} (\bibinfo {year} {2012})}\BibitemShut {NoStop}%
\bibitem [{\citenamefont {Wang}\ \emph {et~al.}(2020)\citenamefont {Wang}, \citenamefont {Sciarrino}, \citenamefont {Laing},\ and\ \citenamefont {Thompson}}]{Wang2020}%
  \BibitemOpen
  \bibfield  {author} {\bibinfo {author} {\bibfnamefont {J.}~\bibnamefont {Wang}}, \bibinfo {author} {\bibfnamefont {F.}~\bibnamefont {Sciarrino}}, \bibinfo {author} {\bibfnamefont {A.}~\bibnamefont {Laing}},\ and\ \bibinfo {author} {\bibfnamefont {M.~G.}\ \bibnamefont {Thompson}},\ }\bibfield  {title} {\bibinfo {title} {Integrated photonic quantum technologies},\ }\href {https://doi.org/10.1038/s41566-019-0532-1} {\bibfield  {journal} {\bibinfo  {journal} {Nature Photonics}\ }\textbf {\bibinfo {volume} {14}},\ \bibinfo {pages} {273} (\bibinfo {year} {2020})}\BibitemShut {NoStop}%
\bibitem [{\citenamefont {Vigliar}\ \emph {et~al.}(2021)\citenamefont {Vigliar}, \citenamefont {Paesani}, \citenamefont {Ding}, \citenamefont {Adcock}, \citenamefont {Wang}, \citenamefont {Morley-Short}, \citenamefont {Bacco}, \citenamefont {Oxenl{\o}we}, \citenamefont {Thompson}, \citenamefont {Rarity},\ and\ \citenamefont {Laing}}]{Vigliar2021}%
  \BibitemOpen
  \bibfield  {author} {\bibinfo {author} {\bibfnamefont {C.}~\bibnamefont {Vigliar}}, \bibinfo {author} {\bibfnamefont {S.}~\bibnamefont {Paesani}}, \bibinfo {author} {\bibfnamefont {Y.}~\bibnamefont {Ding}}, \bibinfo {author} {\bibfnamefont {J.~C.}\ \bibnamefont {Adcock}}, \bibinfo {author} {\bibfnamefont {J.}~\bibnamefont {Wang}}, \bibinfo {author} {\bibfnamefont {S.}~\bibnamefont {Morley-Short}}, \bibinfo {author} {\bibfnamefont {D.}~\bibnamefont {Bacco}}, \bibinfo {author} {\bibfnamefont {L.~K.}\ \bibnamefont {Oxenl{\o}we}}, \bibinfo {author} {\bibfnamefont {M.~G.}\ \bibnamefont {Thompson}}, \bibinfo {author} {\bibfnamefont {J.~G.}\ \bibnamefont {Rarity}},\ and\ \bibinfo {author} {\bibfnamefont {A.}~\bibnamefont {Laing}},\ }\bibfield  {title} {\bibinfo {title} {Error-protected qubits in a silicon photonic chip},\ }\href {https://doi.org/10.1038/s41567-021-01333-w} {\bibfield  {journal} {\bibinfo  {journal} {Nature Physics}\ }\textbf {\bibinfo {volume} {17}},\ \bibinfo {pages} {1137} (\bibinfo {year}
  {2021})}\BibitemShut {NoStop}%
\bibitem [{\citenamefont {Bourassa}\ \emph {et~al.}(2021)\citenamefont {Bourassa}, \citenamefont {Alexander}, \citenamefont {Vasmer}, \citenamefont {Patil}, \citenamefont {Tzitrin}, \citenamefont {Matsuura}, \citenamefont {Su}, \citenamefont {Baragiola}, \citenamefont {Guha}, \citenamefont {Dauphinais}, \citenamefont {Sabapathy}, \citenamefont {Menicucci},\ and\ \citenamefont {Dhand}}]{Bourassa2021}%
  \BibitemOpen
  \bibfield  {author} {\bibinfo {author} {\bibfnamefont {J.~E.}\ \bibnamefont {Bourassa}}, \bibinfo {author} {\bibfnamefont {R.~N.}\ \bibnamefont {Alexander}}, \bibinfo {author} {\bibfnamefont {M.}~\bibnamefont {Vasmer}}, \bibinfo {author} {\bibfnamefont {A.}~\bibnamefont {Patil}}, \bibinfo {author} {\bibfnamefont {I.}~\bibnamefont {Tzitrin}}, \bibinfo {author} {\bibfnamefont {T.}~\bibnamefont {Matsuura}}, \bibinfo {author} {\bibfnamefont {D.}~\bibnamefont {Su}}, \bibinfo {author} {\bibfnamefont {B.~Q.}\ \bibnamefont {Baragiola}}, \bibinfo {author} {\bibfnamefont {S.}~\bibnamefont {Guha}}, \bibinfo {author} {\bibfnamefont {G.}~\bibnamefont {Dauphinais}}, \bibinfo {author} {\bibfnamefont {K.~K.}\ \bibnamefont {Sabapathy}}, \bibinfo {author} {\bibfnamefont {N.~C.}\ \bibnamefont {Menicucci}},\ and\ \bibinfo {author} {\bibfnamefont {I.}~\bibnamefont {Dhand}},\ }\bibfield  {title} {\bibinfo {title} {Blueprint for a {S}calable {P}hotonic {F}ault-{T}olerant {Q}uantum {C}omputer},\ }\href
  {https://doi.org/10.22331/q-2021-02-04-392} {\bibfield  {journal} {\bibinfo  {journal} {{Quantum}}\ }\textbf {\bibinfo {volume} {5}},\ \bibinfo {pages} {392} (\bibinfo {year} {2021})}\BibitemShut {NoStop}%
\bibitem [{\citenamefont {O'Brien}\ \emph {et~al.}(2003)\citenamefont {O'Brien}, \citenamefont {Pryde}, \citenamefont {White}, \citenamefont {Ralph},\ and\ \citenamefont {Branning}}]{O'Brien2003}%
  \BibitemOpen
  \bibfield  {author} {\bibinfo {author} {\bibfnamefont {J.~L.}\ \bibnamefont {O'Brien}}, \bibinfo {author} {\bibfnamefont {G.~J.}\ \bibnamefont {Pryde}}, \bibinfo {author} {\bibfnamefont {A.~G.}\ \bibnamefont {White}}, \bibinfo {author} {\bibfnamefont {T.~C.}\ \bibnamefont {Ralph}},\ and\ \bibinfo {author} {\bibfnamefont {D.}~\bibnamefont {Branning}},\ }\bibfield  {title} {\bibinfo {title} {Demonstration of an all-optical quantum controlled-not gate},\ }\href {https://doi.org/10.1038/nature02054} {\bibfield  {journal} {\bibinfo  {journal} {Nature}\ }\textbf {\bibinfo {volume} {426}},\ \bibinfo {pages} {264} (\bibinfo {year} {2003})}\BibitemShut {NoStop}%
\bibitem [{\citenamefont {Smith}\ \emph {et~al.}(2009)\citenamefont {Smith}, \citenamefont {Kundys}, \citenamefont {Thomas-Peter}, \citenamefont {Smith},\ and\ \citenamefont {Walmsley}}]{Smith2009}%
  \BibitemOpen
  \bibfield  {author} {\bibinfo {author} {\bibfnamefont {B.~J.}\ \bibnamefont {Smith}}, \bibinfo {author} {\bibfnamefont {D.}~\bibnamefont {Kundys}}, \bibinfo {author} {\bibfnamefont {N.}~\bibnamefont {Thomas-Peter}}, \bibinfo {author} {\bibfnamefont {P.~G.~R.}\ \bibnamefont {Smith}},\ and\ \bibinfo {author} {\bibfnamefont {I.~A.}\ \bibnamefont {Walmsley}},\ }\bibfield  {title} {\bibinfo {title} {Phase-controlled integrated photonic quantum circuits},\ }\href {https://doi.org/10.1364/OE.17.013516} {\bibfield  {journal} {\bibinfo  {journal} {Opt. Express}\ }\textbf {\bibinfo {volume} {17}},\ \bibinfo {pages} {13516} (\bibinfo {year} {2009})}\BibitemShut {NoStop}%
\bibitem [{\citenamefont {Crespi}\ \emph {et~al.}(2011)\citenamefont {Crespi}, \citenamefont {Ramponi}, \citenamefont {Osellame}, \citenamefont {Sansoni}, \citenamefont {Bongioanni}, \citenamefont {Sciarrino}, \citenamefont {Vallone},\ and\ \citenamefont {Mataloni}}]{Crespi2011}%
  \BibitemOpen
  \bibfield  {author} {\bibinfo {author} {\bibfnamefont {A.}~\bibnamefont {Crespi}}, \bibinfo {author} {\bibfnamefont {R.}~\bibnamefont {Ramponi}}, \bibinfo {author} {\bibfnamefont {R.}~\bibnamefont {Osellame}}, \bibinfo {author} {\bibfnamefont {L.}~\bibnamefont {Sansoni}}, \bibinfo {author} {\bibfnamefont {I.}~\bibnamefont {Bongioanni}}, \bibinfo {author} {\bibfnamefont {F.}~\bibnamefont {Sciarrino}}, \bibinfo {author} {\bibfnamefont {G.}~\bibnamefont {Vallone}},\ and\ \bibinfo {author} {\bibfnamefont {P.}~\bibnamefont {Mataloni}},\ }\bibfield  {title} {\bibinfo {title} {Integrated photonic quantum gates for polarization qubits},\ }\href {https://doi.org/10.1038/ncomms1570} {\bibfield  {journal} {\bibinfo  {journal} {Nature Communications}\ }\textbf {\bibinfo {volume} {2}},\ \bibinfo {pages} {566} (\bibinfo {year} {2011})}\BibitemShut {NoStop}%
\bibitem [{\citenamefont {Barz}\ \emph {et~al.}(2014)\citenamefont {Barz}, \citenamefont {Kassal}, \citenamefont {Ringbauer}, \citenamefont {Lipp}, \citenamefont {Daki{\'{c}}}, \citenamefont {Aspuru-Guzik},\ and\ \citenamefont {Walther}}]{Barz2014}%
  \BibitemOpen
  \bibfield  {author} {\bibinfo {author} {\bibfnamefont {S.}~\bibnamefont {Barz}}, \bibinfo {author} {\bibfnamefont {I.}~\bibnamefont {Kassal}}, \bibinfo {author} {\bibfnamefont {M.}~\bibnamefont {Ringbauer}}, \bibinfo {author} {\bibfnamefont {Y.~O.}\ \bibnamefont {Lipp}}, \bibinfo {author} {\bibfnamefont {B.}~\bibnamefont {Daki{\'{c}}}}, \bibinfo {author} {\bibfnamefont {A.}~\bibnamefont {Aspuru-Guzik}},\ and\ \bibinfo {author} {\bibfnamefont {P.}~\bibnamefont {Walther}},\ }\bibfield  {title} {\bibinfo {title} {A two-qubit photonic quantum processor and its application to solving systems of linear equations},\ }\href {https://doi.org/10.1038/srep06115} {\bibfield  {journal} {\bibinfo  {journal} {Scientific Reports}\ }\textbf {\bibinfo {volume} {4}},\ \bibinfo {pages} {6115} (\bibinfo {year} {2014})}\BibitemShut {NoStop}%
\bibitem [{\citenamefont {Shi}\ \emph {et~al.}(2022)\citenamefont {Shi}, \citenamefont {Xu}, \citenamefont {Zhang}, \citenamefont {Ye}, \citenamefont {Xiang}, \citenamefont {Liu}, \citenamefont {Wang}, \citenamefont {Su},\ and\ \citenamefont {Li}}]{Shi2022}%
  \BibitemOpen
  \bibfield  {author} {\bibinfo {author} {\bibfnamefont {S.}~\bibnamefont {Shi}}, \bibinfo {author} {\bibfnamefont {B.}~\bibnamefont {Xu}}, \bibinfo {author} {\bibfnamefont {K.}~\bibnamefont {Zhang}}, \bibinfo {author} {\bibfnamefont {G.-S.}\ \bibnamefont {Ye}}, \bibinfo {author} {\bibfnamefont {D.-S.}\ \bibnamefont {Xiang}}, \bibinfo {author} {\bibfnamefont {Y.}~\bibnamefont {Liu}}, \bibinfo {author} {\bibfnamefont {J.}~\bibnamefont {Wang}}, \bibinfo {author} {\bibfnamefont {D.}~\bibnamefont {Su}},\ and\ \bibinfo {author} {\bibfnamefont {L.}~\bibnamefont {Li}},\ }\bibfield  {title} {\bibinfo {title} {High-fidelity photonic quantum logic gate based on near-optimal rydberg single-photon source},\ }\href {https://doi.org/10.1038/s41467-022-32083-9} {\bibfield  {journal} {\bibinfo  {journal} {Nature Communications}\ }\textbf {\bibinfo {volume} {13}},\ \bibinfo {pages} {4454} (\bibinfo {year} {2022})}\BibitemShut {NoStop}%
\bibitem [{\citenamefont {Politi}\ \emph {et~al.}(2009)\citenamefont {Politi}, \citenamefont {Matthews},\ and\ \citenamefont {O'Brien}}]{Politi2009}%
  \BibitemOpen
  \bibfield  {author} {\bibinfo {author} {\bibfnamefont {A.}~\bibnamefont {Politi}}, \bibinfo {author} {\bibfnamefont {J.~C.~F.}\ \bibnamefont {Matthews}},\ and\ \bibinfo {author} {\bibfnamefont {J.~L.}\ \bibnamefont {O'Brien}},\ }\bibfield  {title} {\bibinfo {title} {Shor’s quantum factoring algorithm on a photonic chip},\ }\href {https://doi.org/10.1126/science.1173731} {\bibfield  {journal} {\bibinfo  {journal} {Science}\ }\textbf {\bibinfo {volume} {325}},\ \bibinfo {pages} {1221} (\bibinfo {year} {2009})}\BibitemShut {NoStop}%
\bibitem [{\citenamefont {Bache}\ \emph {et~al.}(2013)\citenamefont {Bache}, \citenamefont {Guo}, \citenamefont {Zhou},\ and\ \citenamefont {Zeng}}]{Bache2013}%
  \BibitemOpen
  \bibfield  {author} {\bibinfo {author} {\bibfnamefont {M.}~\bibnamefont {Bache}}, \bibinfo {author} {\bibfnamefont {H.}~\bibnamefont {Guo}}, \bibinfo {author} {\bibfnamefont {B.}~\bibnamefont {Zhou}},\ and\ \bibinfo {author} {\bibfnamefont {X.}~\bibnamefont {Zeng}},\ }\bibfield  {title} {\bibinfo {title} {The anisotropic kerr nonlinear refractive index of the beta-barium borate ($\beta$-bab2o4) nonlinear crystal},\ }\href {https://doi.org/10.1364/OME.3.000357} {\bibfield  {journal} {\bibinfo  {journal} {Opt. Mater. Express}\ }\textbf {\bibinfo {volume} {3}},\ \bibinfo {pages} {357} (\bibinfo {year} {2013})}\BibitemShut {NoStop}%
\bibitem [{\citenamefont {Li}\ \emph {et~al.}(1997)\citenamefont {Li}, \citenamefont {Zhou}, \citenamefont {Zhang},\ and\ \citenamefont {Ji}}]{Li1997}%
  \BibitemOpen
  \bibfield  {author} {\bibinfo {author} {\bibfnamefont {H.}~\bibnamefont {Li}}, \bibinfo {author} {\bibfnamefont {F.}~\bibnamefont {Zhou}}, \bibinfo {author} {\bibfnamefont {X.}~\bibnamefont {Zhang}},\ and\ \bibinfo {author} {\bibfnamefont {W.}~\bibnamefont {Ji}},\ }\bibfield  {title} {\bibinfo {title} {Picosecond z-scan study of bound electronic kerr effect in linbo3 crystal associated with two-photon absorption},\ }\href {https://doi.org/10.1007/s003400050229} {\bibfield  {journal} {\bibinfo  {journal} {Applied Physics B}\ }\textbf {\bibinfo {volume} {64}},\ \bibinfo {pages} {659} (\bibinfo {year} {1997})}\BibitemShut {NoStop}%
\bibitem [{\citenamefont {Ludlow}\ \emph {et~al.}(2001)\citenamefont {Ludlow}, \citenamefont {Nelson},\ and\ \citenamefont {Bergeson}}]{Ludlow2001}%
  \BibitemOpen
  \bibfield  {author} {\bibinfo {author} {\bibfnamefont {A.~D.}\ \bibnamefont {Ludlow}}, \bibinfo {author} {\bibfnamefont {H.~M.}\ \bibnamefont {Nelson}},\ and\ \bibinfo {author} {\bibfnamefont {S.~D.}\ \bibnamefont {Bergeson}},\ }\bibfield  {title} {\bibinfo {title} {Two-photon absorption in potassium niobate},\ }\href@noop {} {\bibfield  {journal} {\bibinfo  {journal} {Journal of The Optical Society of America B-optical Physics}\ }\textbf {\bibinfo {volume} {18}},\ \bibinfo {pages} {1813} (\bibinfo {year} {2001})}\BibitemShut {NoStop}%
\bibitem [{\citenamefont {Moreno-Cardoner}\ \emph {et~al.}(2021)\citenamefont {Moreno-Cardoner}, \citenamefont {Goncalves},\ and\ \citenamefont {Chang}}]{Moreno2021}%
  \BibitemOpen
  \bibfield  {author} {\bibinfo {author} {\bibfnamefont {M.}~\bibnamefont {Moreno-Cardoner}}, \bibinfo {author} {\bibfnamefont {D.}~\bibnamefont {Goncalves}},\ and\ \bibinfo {author} {\bibfnamefont {D.~E.}\ \bibnamefont {Chang}},\ }\bibfield  {title} {\bibinfo {title} {Quantum nonlinear optics based on two-dimensional rydberg atom arrays},\ }\href {https://doi.org/10.1103/PhysRevLett.127.263602} {\bibfield  {journal} {\bibinfo  {journal} {Phys. Rev. Lett.}\ }\textbf {\bibinfo {volume} {127}},\ \bibinfo {pages} {263602} (\bibinfo {year} {2021})}\BibitemShut {NoStop}%
\bibitem [{\citenamefont {Mu}\ \emph {et~al.}(2021)\citenamefont {Mu}, \citenamefont {Qin}, \citenamefont {Shi},\ and\ \citenamefont {Huang}}]{Mu2021}%
  \BibitemOpen
  \bibfield  {author} {\bibinfo {author} {\bibfnamefont {Y.}~\bibnamefont {Mu}}, \bibinfo {author} {\bibfnamefont {L.}~\bibnamefont {Qin}}, \bibinfo {author} {\bibfnamefont {Z.}~\bibnamefont {Shi}},\ and\ \bibinfo {author} {\bibfnamefont {G.}~\bibnamefont {Huang}},\ }\bibfield  {title} {\bibinfo {title} {Giant kerr nonlinearities and magneto-optical rotations in a rydberg-atom gas via double electromagnetically induced transparency},\ }\href {https://doi.org/10.1103/PhysRevA.103.043709} {\bibfield  {journal} {\bibinfo  {journal} {Phys. Rev. A}\ }\textbf {\bibinfo {volume} {103}},\ \bibinfo {pages} {043709} (\bibinfo {year} {2021})}\BibitemShut {NoStop}%
\bibitem [{\citenamefont {Chen}\ \emph {et~al.}(2021)\citenamefont {Chen}, \citenamefont {Yang}, \citenamefont {Wu}, \citenamefont {Shen}, \citenamefont {Tey},\ and\ \citenamefont {You}}]{Chen2021}%
  \BibitemOpen
  \bibfield  {author} {\bibinfo {author} {\bibfnamefont {C.}~\bibnamefont {Chen}}, \bibinfo {author} {\bibfnamefont {F.}~\bibnamefont {Yang}}, \bibinfo {author} {\bibfnamefont {X.}~\bibnamefont {Wu}}, \bibinfo {author} {\bibfnamefont {C.}~\bibnamefont {Shen}}, \bibinfo {author} {\bibfnamefont {M.~K.}\ \bibnamefont {Tey}},\ and\ \bibinfo {author} {\bibfnamefont {L.}~\bibnamefont {You}},\ }\bibfield  {title} {\bibinfo {title} {Two-color optical nonlinearity in an ultracold rydberg atom gas mixture},\ }\href {https://doi.org/10.1103/PhysRevA.103.053303} {\bibfield  {journal} {\bibinfo  {journal} {Phys. Rev. A}\ }\textbf {\bibinfo {volume} {103}},\ \bibinfo {pages} {053303} (\bibinfo {year} {2021})}\BibitemShut {NoStop}%
\bibitem [{\citenamefont {Bai}\ and\ \citenamefont {Huang}(2016)}]{Bai:16}%
  \BibitemOpen
  \bibfield  {author} {\bibinfo {author} {\bibfnamefont {Z.}~\bibnamefont {Bai}}\ and\ \bibinfo {author} {\bibfnamefont {G.}~\bibnamefont {Huang}},\ }\bibfield  {title} {\bibinfo {title} {Enhanced third-order and fifth-order kerr nonlinearities in a cold atomic system via rydberg-rydberg interaction},\ }\href {https://doi.org/10.1364/OE.24.004442} {\bibfield  {journal} {\bibinfo  {journal} {Opt. Express}\ }\textbf {\bibinfo {volume} {24}},\ \bibinfo {pages} {4442} (\bibinfo {year} {2016})}\BibitemShut {NoStop}%
\bibitem [{\citenamefont {Hermann-Avigliano}\ \emph {et~al.}(2014)\citenamefont {Hermann-Avigliano}, \citenamefont {Teixeira}, \citenamefont {Nguyen}, \citenamefont {Cantat-Moltrecht}, \citenamefont {Nogues}, \citenamefont {Dotsenko}, \citenamefont {Gleyzes}, \citenamefont {Raimond}, \citenamefont {Haroche},\ and\ \citenamefont {Brune}}]{Hermann2014}%
  \BibitemOpen
  \bibfield  {author} {\bibinfo {author} {\bibfnamefont {C.}~\bibnamefont {Hermann-Avigliano}}, \bibinfo {author} {\bibfnamefont {R.~C.}\ \bibnamefont {Teixeira}}, \bibinfo {author} {\bibfnamefont {T.~L.}\ \bibnamefont {Nguyen}}, \bibinfo {author} {\bibfnamefont {T.}~\bibnamefont {Cantat-Moltrecht}}, \bibinfo {author} {\bibfnamefont {G.}~\bibnamefont {Nogues}}, \bibinfo {author} {\bibfnamefont {I.}~\bibnamefont {Dotsenko}}, \bibinfo {author} {\bibfnamefont {S.}~\bibnamefont {Gleyzes}}, \bibinfo {author} {\bibfnamefont {J.~M.}\ \bibnamefont {Raimond}}, \bibinfo {author} {\bibfnamefont {S.}~\bibnamefont {Haroche}},\ and\ \bibinfo {author} {\bibfnamefont {M.}~\bibnamefont {Brune}},\ }\bibfield  {title} {\bibinfo {title} {Long coherence times for rydberg qubits on a superconducting atom chip},\ }\href {https://doi.org/10.1103/PhysRevA.90.040502} {\bibfield  {journal} {\bibinfo  {journal} {Phys. Rev. A}\ }\textbf {\bibinfo {volume} {90}},\ \bibinfo {pages} {040502} (\bibinfo {year} {2014})}\BibitemShut {NoStop}%
\bibitem [{\citenamefont {Lampen}\ \emph {et~al.}(2018)\citenamefont {Lampen}, \citenamefont {Nguyen}, \citenamefont {Li}, \citenamefont {Berman},\ and\ \citenamefont {Kuzmich}}]{Lampen2018}%
  \BibitemOpen
  \bibfield  {author} {\bibinfo {author} {\bibfnamefont {J.}~\bibnamefont {Lampen}}, \bibinfo {author} {\bibfnamefont {H.}~\bibnamefont {Nguyen}}, \bibinfo {author} {\bibfnamefont {L.}~\bibnamefont {Li}}, \bibinfo {author} {\bibfnamefont {P.~R.}\ \bibnamefont {Berman}},\ and\ \bibinfo {author} {\bibfnamefont {A.}~\bibnamefont {Kuzmich}},\ }\bibfield  {title} {\bibinfo {title} {Long-lived coherence between ground and rydberg levels in a magic-wavelength lattice},\ }\href {https://doi.org/10.1103/PhysRevA.98.033411} {\bibfield  {journal} {\bibinfo  {journal} {Phys. Rev. A}\ }\textbf {\bibinfo {volume} {98}},\ \bibinfo {pages} {033411} (\bibinfo {year} {2018})}\BibitemShut {NoStop}%
\bibitem [{\citenamefont {Cantat-Moltrecht}\ \emph {et~al.}(2020)\citenamefont {Cantat-Moltrecht}, \citenamefont {Corti\~nas}, \citenamefont {Ravon}, \citenamefont {M\'ehaignerie}, \citenamefont {Haroche}, \citenamefont {Raimond}, \citenamefont {Favier}, \citenamefont {Brune},\ and\ \citenamefont {Sayrin}}]{Contat2020}%
  \BibitemOpen
  \bibfield  {author} {\bibinfo {author} {\bibfnamefont {T.}~\bibnamefont {Cantat-Moltrecht}}, \bibinfo {author} {\bibfnamefont {R.}~\bibnamefont {Corti\~nas}}, \bibinfo {author} {\bibfnamefont {B.}~\bibnamefont {Ravon}}, \bibinfo {author} {\bibfnamefont {P.}~\bibnamefont {M\'ehaignerie}}, \bibinfo {author} {\bibfnamefont {S.}~\bibnamefont {Haroche}}, \bibinfo {author} {\bibfnamefont {J.~M.}\ \bibnamefont {Raimond}}, \bibinfo {author} {\bibfnamefont {M.}~\bibnamefont {Favier}}, \bibinfo {author} {\bibfnamefont {M.}~\bibnamefont {Brune}},\ and\ \bibinfo {author} {\bibfnamefont {C.}~\bibnamefont {Sayrin}},\ }\bibfield  {title} {\bibinfo {title} {Long-lived circular rydberg states of laser-cooled rubidium atoms in a cryostat},\ }\href {https://doi.org/10.1103/PhysRevResearch.2.022032} {\bibfield  {journal} {\bibinfo  {journal} {Phys. Rev. Res.}\ }\textbf {\bibinfo {volume} {2}},\ \bibinfo {pages} {022032} (\bibinfo {year} {2020})}\BibitemShut {NoStop}%
\bibitem [{\citenamefont {Gallagher}(1994)}]{Gallagher1994}%
  \BibitemOpen
  \bibfield  {author} {\bibinfo {author} {\bibfnamefont {T.~F.}\ \bibnamefont {Gallagher}},\ }\href@noop {} {\emph {\bibinfo {title} {Rydberg Atoms}}},\ Cambridge Monographs on Atomic, Molecular and Chemical Physics\ (\bibinfo  {publisher} {Cambridge University Press},\ \bibinfo {year} {1994})\BibitemShut {NoStop}%
\bibitem [{\citenamefont {Adams}\ \emph {et~al.}(2019)\citenamefont {Adams}, \citenamefont {Pritchard},\ and\ \citenamefont {Shaffer}}]{Adams2020}%
  \BibitemOpen
  \bibfield  {author} {\bibinfo {author} {\bibfnamefont {C.~S.}\ \bibnamefont {Adams}}, \bibinfo {author} {\bibfnamefont {J.~D.}\ \bibnamefont {Pritchard}},\ and\ \bibinfo {author} {\bibfnamefont {J.~P.}\ \bibnamefont {Shaffer}},\ }\bibfield  {title} {\bibinfo {title} {Rydberg atom quantum technologies},\ }\href {https://doi.org/10.1088/1361-6455/ab52ef} {\bibfield  {journal} {\bibinfo  {journal} {Journal of Physics B: Atomic, Molecular and Optical Physics}\ }\textbf {\bibinfo {volume} {53}},\ \bibinfo {pages} {012002} (\bibinfo {year} {2019})}\BibitemShut {NoStop}%
\bibitem [{\citenamefont {Jiao}\ \emph {et~al.}(2022)\citenamefont {Jiao}, \citenamefont {Bai}, \citenamefont {Song}, \citenamefont {Bao}, \citenamefont {Zhao},\ and\ \citenamefont {Jia}}]{Jiao2022}%
  \BibitemOpen
  \bibfield  {author} {\bibinfo {author} {\bibfnamefont {Y.}~\bibnamefont {Jiao}}, \bibinfo {author} {\bibfnamefont {J.}~\bibnamefont {Bai}}, \bibinfo {author} {\bibfnamefont {R.}~\bibnamefont {Song}}, \bibinfo {author} {\bibfnamefont {S.}~\bibnamefont {Bao}}, \bibinfo {author} {\bibfnamefont {J.}~\bibnamefont {Zhao}},\ and\ \bibinfo {author} {\bibfnamefont {S.}~\bibnamefont {Jia}},\ }\bibfield  {title} {\bibinfo {title} {Electric field tuned dipolar interaction between rydberg atoms},\ }\bibfield  {journal} {\bibinfo  {journal} {Frontiers in Physics}\ }\textbf {\bibinfo {volume} {10}},\ \href {https://doi.org/10.3389/fphy.2022.892542} {10.3389/fphy.2022.892542} (\bibinfo {year} {2022})\BibitemShut {NoStop}%
\bibitem [{\citenamefont {Saffman}\ \emph {et~al.}(2010)\citenamefont {Saffman}, \citenamefont {Walker},\ and\ \citenamefont {M\o{}lmer}}]{Saffman2010}%
  \BibitemOpen
  \bibfield  {author} {\bibinfo {author} {\bibfnamefont {M.}~\bibnamefont {Saffman}}, \bibinfo {author} {\bibfnamefont {T.~G.}\ \bibnamefont {Walker}},\ and\ \bibinfo {author} {\bibfnamefont {K.}~\bibnamefont {M\o{}lmer}},\ }\bibfield  {title} {\bibinfo {title} {Quantum information with rydberg atoms},\ }\href {https://doi.org/10.1103/RevModPhys.82.2313} {\bibfield  {journal} {\bibinfo  {journal} {Rev. Mod. Phys.}\ }\textbf {\bibinfo {volume} {82}},\ \bibinfo {pages} {2313} (\bibinfo {year} {2010})}\BibitemShut {NoStop}%
\bibitem [{\citenamefont {Peyronel}\ \emph {et~al.}(2012)\citenamefont {Peyronel}, \citenamefont {Firstenberg}, \citenamefont {Liang}, \citenamefont {Hofferberth}, \citenamefont {Gorshkov}, \citenamefont {Pohl}, \citenamefont {Lukin},\ and\ \citenamefont {Vuleti{\'{c}}}}]{Peyronel2012}%
  \BibitemOpen
  \bibfield  {author} {\bibinfo {author} {\bibfnamefont {T.}~\bibnamefont {Peyronel}}, \bibinfo {author} {\bibfnamefont {O.}~\bibnamefont {Firstenberg}}, \bibinfo {author} {\bibfnamefont {Q.-Y.}\ \bibnamefont {Liang}}, \bibinfo {author} {\bibfnamefont {S.}~\bibnamefont {Hofferberth}}, \bibinfo {author} {\bibfnamefont {A.~V.}\ \bibnamefont {Gorshkov}}, \bibinfo {author} {\bibfnamefont {T.}~\bibnamefont {Pohl}}, \bibinfo {author} {\bibfnamefont {M.~D.}\ \bibnamefont {Lukin}},\ and\ \bibinfo {author} {\bibfnamefont {V.}~\bibnamefont {Vuleti{\'{c}}}},\ }\bibfield  {title} {\bibinfo {title} {Quantum nonlinear optics with single photons enabled by strongly interacting atoms},\ }\href {https://doi.org/10.1038/nature11361} {\bibfield  {journal} {\bibinfo  {journal} {Nature}\ }\textbf {\bibinfo {volume} {488}},\ \bibinfo {pages} {57} (\bibinfo {year} {2012})}\BibitemShut {NoStop}%
\bibitem [{\citenamefont {Ripka}\ \emph {et~al.}(2018)\citenamefont {Ripka}, \citenamefont {Kübler}, \citenamefont {Löw},\ and\ \citenamefont {Pfau}}]{Ripka2018}%
  \BibitemOpen
  \bibfield  {author} {\bibinfo {author} {\bibfnamefont {F.}~\bibnamefont {Ripka}}, \bibinfo {author} {\bibfnamefont {H.}~\bibnamefont {Kübler}}, \bibinfo {author} {\bibfnamefont {R.}~\bibnamefont {Löw}},\ and\ \bibinfo {author} {\bibfnamefont {T.}~\bibnamefont {Pfau}},\ }\bibfield  {title} {\bibinfo {title} {A room-temperature single-photon source based on strongly interacting rydberg atoms},\ }\href {https://doi.org/10.1126/science.aau1949} {\bibfield  {journal} {\bibinfo  {journal} {Science}\ }\textbf {\bibinfo {volume} {362}},\ \bibinfo {pages} {446} (\bibinfo {year} {2018})}\BibitemShut {NoStop}%
\bibitem [{\citenamefont {Kumlin}\ \emph {et~al.}(2023)\citenamefont {Kumlin}, \citenamefont {Braun}, \citenamefont {Tresp}, \citenamefont {Stiesdal}, \citenamefont {Hofferberth},\ and\ \citenamefont {Paris-Mandoki}}]{Kumlin_2023}%
  \BibitemOpen
  \bibfield  {author} {\bibinfo {author} {\bibfnamefont {J.}~\bibnamefont {Kumlin}}, \bibinfo {author} {\bibfnamefont {C.}~\bibnamefont {Braun}}, \bibinfo {author} {\bibfnamefont {C.}~\bibnamefont {Tresp}}, \bibinfo {author} {\bibfnamefont {N.}~\bibnamefont {Stiesdal}}, \bibinfo {author} {\bibfnamefont {S.}~\bibnamefont {Hofferberth}},\ and\ \bibinfo {author} {\bibfnamefont {A.}~\bibnamefont {Paris-Mandoki}},\ }\bibfield  {title} {\bibinfo {title} {Quantum optics with rydberg superatoms},\ }\href {https://doi.org/10.1088/2399-6528/acd51d} {\bibfield  {journal} {\bibinfo  {journal} {Journal of Physics Communications}\ }\textbf {\bibinfo {volume} {7}},\ \bibinfo {pages} {052001} (\bibinfo {year} {2023})}\BibitemShut {NoStop}%
\bibitem [{\citenamefont {Bernien}\ \emph {et~al.}(2017)\citenamefont {Bernien}, \citenamefont {Schwartz}, \citenamefont {Keesling}, \citenamefont {Levine}, \citenamefont {Omran}, \citenamefont {Pichler}, \citenamefont {Choi}, \citenamefont {Zibrov}, \citenamefont {Endres}, \citenamefont {Greiner}, \citenamefont {Vuletić},\ and\ \citenamefont {Lukin}}]{Bernien2017}%
  \BibitemOpen
  \bibfield  {author} {\bibinfo {author} {\bibfnamefont {H.}~\bibnamefont {Bernien}}, \bibinfo {author} {\bibfnamefont {S.}~\bibnamefont {Schwartz}}, \bibinfo {author} {\bibfnamefont {A.}~\bibnamefont {Keesling}}, \bibinfo {author} {\bibfnamefont {H.}~\bibnamefont {Levine}}, \bibinfo {author} {\bibfnamefont {A.}~\bibnamefont {Omran}}, \bibinfo {author} {\bibfnamefont {H.}~\bibnamefont {Pichler}}, \bibinfo {author} {\bibfnamefont {S.}~\bibnamefont {Choi}}, \bibinfo {author} {\bibfnamefont {A.~S.}\ \bibnamefont {Zibrov}}, \bibinfo {author} {\bibfnamefont {M.}~\bibnamefont {Endres}}, \bibinfo {author} {\bibfnamefont {M.}~\bibnamefont {Greiner}}, \bibinfo {author} {\bibfnamefont {V.}~\bibnamefont {Vuletić}},\ and\ \bibinfo {author} {\bibfnamefont {M.~D.}\ \bibnamefont {Lukin}},\ }\bibfield  {title} {\bibinfo {title} {Probing many-body dynamics on a 51-atom quantum simulator},\ }\href {https://doi.org/10.1038/nature24622} {\bibfield  {journal} {\bibinfo  {journal} {Nature}\ }\textbf {\bibinfo {volume} {551}},\
  \bibinfo {pages} {579–584} (\bibinfo {year} {2017})}\BibitemShut {NoStop}%
\bibitem [{\citenamefont {Scholl}\ \emph {et~al.}(2021)\citenamefont {Scholl}, \citenamefont {Schuler}, \citenamefont {Williams}, \citenamefont {Eberharter}, \citenamefont {Barredo}, \citenamefont {Schymik}, \citenamefont {Lienhard}, \citenamefont {Henry}, \citenamefont {Lang}, \citenamefont {Lahaye}, \citenamefont {L\"{a}uchli},\ and\ \citenamefont {Browaeys}}]{Scholl2021}%
  \BibitemOpen
  \bibfield  {author} {\bibinfo {author} {\bibfnamefont {P.}~\bibnamefont {Scholl}}, \bibinfo {author} {\bibfnamefont {M.}~\bibnamefont {Schuler}}, \bibinfo {author} {\bibfnamefont {H.~J.}\ \bibnamefont {Williams}}, \bibinfo {author} {\bibfnamefont {A.~A.}\ \bibnamefont {Eberharter}}, \bibinfo {author} {\bibfnamefont {D.}~\bibnamefont {Barredo}}, \bibinfo {author} {\bibfnamefont {K.-N.}\ \bibnamefont {Schymik}}, \bibinfo {author} {\bibfnamefont {V.}~\bibnamefont {Lienhard}}, \bibinfo {author} {\bibfnamefont {L.-P.}\ \bibnamefont {Henry}}, \bibinfo {author} {\bibfnamefont {T.~C.}\ \bibnamefont {Lang}}, \bibinfo {author} {\bibfnamefont {T.}~\bibnamefont {Lahaye}}, \bibinfo {author} {\bibfnamefont {A.~M.}\ \bibnamefont {L\"{a}uchli}},\ and\ \bibinfo {author} {\bibfnamefont {A.}~\bibnamefont {Browaeys}},\ }\bibfield  {title} {\bibinfo {title} {Quantum simulation of 2d antiferromagnets with hundreds of rydberg atoms},\ }\href {https://doi.org/10.1038/s41586-021-03585-1} {\bibfield  {journal} {\bibinfo  {journal}
  {Nature}\ }\textbf {\bibinfo {volume} {595}},\ \bibinfo {pages} {233–238} (\bibinfo {year} {2021})}\BibitemShut {NoStop}%
\bibitem [{\citenamefont {Bluvstein}\ \emph {et~al.}(2022)\citenamefont {Bluvstein}, \citenamefont {Levine}, \citenamefont {Semeghini}, \citenamefont {Wang}, \citenamefont {Ebadi}, \citenamefont {Kalinowski}, \citenamefont {Keesling}, \citenamefont {Maskara}, \citenamefont {Pichler}, \citenamefont {Greiner}, \citenamefont {Vuleti{\'{c}}},\ and\ \citenamefont {Lukin}}]{Bluvstein2022}%
  \BibitemOpen
  \bibfield  {author} {\bibinfo {author} {\bibfnamefont {D.}~\bibnamefont {Bluvstein}}, \bibinfo {author} {\bibfnamefont {H.}~\bibnamefont {Levine}}, \bibinfo {author} {\bibfnamefont {G.}~\bibnamefont {Semeghini}}, \bibinfo {author} {\bibfnamefont {T.~T.}\ \bibnamefont {Wang}}, \bibinfo {author} {\bibfnamefont {S.}~\bibnamefont {Ebadi}}, \bibinfo {author} {\bibfnamefont {M.}~\bibnamefont {Kalinowski}}, \bibinfo {author} {\bibfnamefont {A.}~\bibnamefont {Keesling}}, \bibinfo {author} {\bibfnamefont {N.}~\bibnamefont {Maskara}}, \bibinfo {author} {\bibfnamefont {H.}~\bibnamefont {Pichler}}, \bibinfo {author} {\bibfnamefont {M.}~\bibnamefont {Greiner}}, \bibinfo {author} {\bibfnamefont {V.}~\bibnamefont {Vuleti{\'{c}}}},\ and\ \bibinfo {author} {\bibfnamefont {M.~D.}\ \bibnamefont {Lukin}},\ }\bibfield  {title} {\bibinfo {title} {A quantum processor based on coherent transport of entangled atom arrays},\ }\href {https://doi.org/10.1038/s41586-022-04592-6} {\bibfield  {journal} {\bibinfo  {journal} {Nature}\
  }\textbf {\bibinfo {volume} {604}},\ \bibinfo {pages} {451} (\bibinfo {year} {2022})}\BibitemShut {NoStop}%
\bibitem [{\citenamefont {Giudici}\ \emph {et~al.}(2022)\citenamefont {Giudici}, \citenamefont {Lukin},\ and\ \citenamefont {Pichler}}]{Giudici2022}%
  \BibitemOpen
  \bibfield  {author} {\bibinfo {author} {\bibfnamefont {G.}~\bibnamefont {Giudici}}, \bibinfo {author} {\bibfnamefont {M.~D.}\ \bibnamefont {Lukin}},\ and\ \bibinfo {author} {\bibfnamefont {H.}~\bibnamefont {Pichler}},\ }\bibfield  {title} {\bibinfo {title} {Dynamical preparation of quantum spin liquids in rydberg atom arrays},\ }\href {https://doi.org/10.1103/PhysRevLett.129.090401} {\bibfield  {journal} {\bibinfo  {journal} {Phys. Rev. Lett.}\ }\textbf {\bibinfo {volume} {129}},\ \bibinfo {pages} {090401} (\bibinfo {year} {2022})}\BibitemShut {NoStop}%
\bibitem [{\citenamefont {Nishad}\ \emph {et~al.}(2023)\citenamefont {Nishad}, \citenamefont {Keselman}, \citenamefont {Lahaye}, \citenamefont {Browaeys},\ and\ \citenamefont {Tsesses}}]{Nishad2023}%
  \BibitemOpen
  \bibfield  {author} {\bibinfo {author} {\bibfnamefont {N.}~\bibnamefont {Nishad}}, \bibinfo {author} {\bibfnamefont {A.}~\bibnamefont {Keselman}}, \bibinfo {author} {\bibfnamefont {T.}~\bibnamefont {Lahaye}}, \bibinfo {author} {\bibfnamefont {A.}~\bibnamefont {Browaeys}},\ and\ \bibinfo {author} {\bibfnamefont {S.}~\bibnamefont {Tsesses}},\ }\bibfield  {title} {\bibinfo {title} {Quantum simulation of generic spin-exchange models in floquet-engineered rydberg-atom arrays},\ }\href {https://doi.org/10.1103/PhysRevA.108.053318} {\bibfield  {journal} {\bibinfo  {journal} {Phys. Rev. A}\ }\textbf {\bibinfo {volume} {108}},\ \bibinfo {pages} {053318} (\bibinfo {year} {2023})}\BibitemShut {NoStop}%
\bibitem [{\citenamefont {Moss}\ \emph {et~al.}(2024)\citenamefont {Moss}, \citenamefont {Ebadi}, \citenamefont {Wang}, \citenamefont {Semeghini}, \citenamefont {Bohrdt}, \citenamefont {Lukin},\ and\ \citenamefont {Melko}}]{Moss2024}%
  \BibitemOpen
  \bibfield  {author} {\bibinfo {author} {\bibfnamefont {M.~S.}\ \bibnamefont {Moss}}, \bibinfo {author} {\bibfnamefont {S.}~\bibnamefont {Ebadi}}, \bibinfo {author} {\bibfnamefont {T.~T.}\ \bibnamefont {Wang}}, \bibinfo {author} {\bibfnamefont {G.}~\bibnamefont {Semeghini}}, \bibinfo {author} {\bibfnamefont {A.}~\bibnamefont {Bohrdt}}, \bibinfo {author} {\bibfnamefont {M.~D.}\ \bibnamefont {Lukin}},\ and\ \bibinfo {author} {\bibfnamefont {R.~G.}\ \bibnamefont {Melko}},\ }\bibfield  {title} {\bibinfo {title} {Enhancing variational monte carlo simulations using a programmable quantum simulator},\ }\href {https://doi.org/10.1103/PhysRevA.109.032410} {\bibfield  {journal} {\bibinfo  {journal} {Phys. Rev. A}\ }\textbf {\bibinfo {volume} {109}},\ \bibinfo {pages} {032410} (\bibinfo {year} {2024})}\BibitemShut {NoStop}%
\bibitem [{\citenamefont {Bluvstein}\ \emph {et~al.}(2024)\citenamefont {Bluvstein}, \citenamefont {Evered}, \citenamefont {Geim}, \citenamefont {Li}, \citenamefont {Zhou}, \citenamefont {Manovitz}, \citenamefont {Ebadi}, \citenamefont {Cain}, \citenamefont {Kalinowski}, \citenamefont {Hangleiter}, \citenamefont {Bonilla~Ataides}, \citenamefont {Maskara}, \citenamefont {Cong}, \citenamefont {Gao}, \citenamefont {Sales~Rodriguez}, \citenamefont {Karolyshyn}, \citenamefont {Semeghini}, \citenamefont {Gullans}, \citenamefont {Greiner}, \citenamefont {Vuleti{\'{c}}},\ and\ \citenamefont {Lukin}}]{Bluvstein2024}%
  \BibitemOpen
  \bibfield  {author} {\bibinfo {author} {\bibfnamefont {D.}~\bibnamefont {Bluvstein}}, \bibinfo {author} {\bibfnamefont {S.~J.}\ \bibnamefont {Evered}}, \bibinfo {author} {\bibfnamefont {A.~A.}\ \bibnamefont {Geim}}, \bibinfo {author} {\bibfnamefont {S.~H.}\ \bibnamefont {Li}}, \bibinfo {author} {\bibfnamefont {H.}~\bibnamefont {Zhou}}, \bibinfo {author} {\bibfnamefont {T.}~\bibnamefont {Manovitz}}, \bibinfo {author} {\bibfnamefont {S.}~\bibnamefont {Ebadi}}, \bibinfo {author} {\bibfnamefont {M.}~\bibnamefont {Cain}}, \bibinfo {author} {\bibfnamefont {M.}~\bibnamefont {Kalinowski}}, \bibinfo {author} {\bibfnamefont {D.}~\bibnamefont {Hangleiter}}, \bibinfo {author} {\bibfnamefont {J.~P.}\ \bibnamefont {Bonilla~Ataides}}, \bibinfo {author} {\bibfnamefont {N.}~\bibnamefont {Maskara}}, \bibinfo {author} {\bibfnamefont {I.}~\bibnamefont {Cong}}, \bibinfo {author} {\bibfnamefont {X.}~\bibnamefont {Gao}}, \bibinfo {author} {\bibfnamefont {P.}~\bibnamefont {Sales~Rodriguez}}, \bibinfo {author} {\bibfnamefont
  {T.}~\bibnamefont {Karolyshyn}}, \bibinfo {author} {\bibfnamefont {G.}~\bibnamefont {Semeghini}}, \bibinfo {author} {\bibfnamefont {M.~J.}\ \bibnamefont {Gullans}}, \bibinfo {author} {\bibfnamefont {M.}~\bibnamefont {Greiner}}, \bibinfo {author} {\bibfnamefont {V.}~\bibnamefont {Vuleti{\'{c}}}},\ and\ \bibinfo {author} {\bibfnamefont {M.~D.}\ \bibnamefont {Lukin}},\ }\bibfield  {title} {\bibinfo {title} {Logical quantum processor based on reconfigurable atom arrays},\ }\href {https://doi.org/10.1038/s41586-023-06927-3} {\bibfield  {journal} {\bibinfo  {journal} {Nature}\ }\textbf {\bibinfo {volume} {626}},\ \bibinfo {pages} {58} (\bibinfo {year} {2024})}\BibitemShut {NoStop}%
\bibitem [{\citenamefont {Michel}\ \emph {et~al.}(2024)\citenamefont {Michel}, \citenamefont {Henriet}, \citenamefont {Domain}, \citenamefont {Browaeys},\ and\ \citenamefont {Ayral}}]{Michel2024}%
  \BibitemOpen
  \bibfield  {author} {\bibinfo {author} {\bibfnamefont {A.}~\bibnamefont {Michel}}, \bibinfo {author} {\bibfnamefont {L.}~\bibnamefont {Henriet}}, \bibinfo {author} {\bibfnamefont {C.}~\bibnamefont {Domain}}, \bibinfo {author} {\bibfnamefont {A.}~\bibnamefont {Browaeys}},\ and\ \bibinfo {author} {\bibfnamefont {T.}~\bibnamefont {Ayral}},\ }\bibfield  {title} {\bibinfo {title} {Hubbard physics with rydberg atoms: Using a quantum spin simulator to simulate strong fermionic correlations},\ }\href {https://doi.org/10.1103/PhysRevB.109.174409} {\bibfield  {journal} {\bibinfo  {journal} {Phys. Rev. B}\ }\textbf {\bibinfo {volume} {109}},\ \bibinfo {pages} {174409} (\bibinfo {year} {2024})}\BibitemShut {NoStop}%
\bibitem [{\citenamefont {Larrouy}\ \emph {et~al.}(2019)\citenamefont {Larrouy}, \citenamefont {Dietsche}, \citenamefont {Richaud}, \citenamefont {Raimond}, \citenamefont {Brune},\ and\ \citenamefont {Gleyzes}}]{Larrouy2019}%
  \BibitemOpen
  \bibfield  {author} {\bibinfo {author} {\bibfnamefont {A.}~\bibnamefont {Larrouy}}, \bibinfo {author} {\bibfnamefont {E.~K.}\ \bibnamefont {Dietsche}}, \bibinfo {author} {\bibfnamefont {R.}~\bibnamefont {Richaud}}, \bibinfo {author} {\bibfnamefont {J.~M.}\ \bibnamefont {Raimond}}, \bibinfo {author} {\bibfnamefont {M.}~\bibnamefont {Brune}},\ and\ \bibinfo {author} {\bibfnamefont {S.}~\bibnamefont {Gleyzes}},\ }\bibfield  {title} {\bibinfo {title} {Quantum sensing using rydberg atoms},\ }in\ \href {https://doi.org/10.1364/QIM.2019.S3A.5} {\emph {\bibinfo {booktitle} {Quantum Information and Measurement (QIM) V: Quantum Technologies}}}\ (\bibinfo  {publisher} {Optica Publishing Group},\ \bibinfo {year} {2019})\ p.\ \bibinfo {pages} {S3A.5}\BibitemShut {NoStop}%
\bibitem [{\citenamefont {Simons}\ \emph {et~al.}(2021{\natexlab{a}})\citenamefont {Simons}, \citenamefont {Artusio-Glimpse}, \citenamefont {Robinson}, \citenamefont {Prajapati},\ and\ \citenamefont {Holloway}}]{Simons2021}%
  \BibitemOpen
  \bibfield  {author} {\bibinfo {author} {\bibfnamefont {M.~T.}\ \bibnamefont {Simons}}, \bibinfo {author} {\bibfnamefont {A.~B.}\ \bibnamefont {Artusio-Glimpse}}, \bibinfo {author} {\bibfnamefont {A.~K.}\ \bibnamefont {Robinson}}, \bibinfo {author} {\bibfnamefont {N.}~\bibnamefont {Prajapati}},\ and\ \bibinfo {author} {\bibfnamefont {C.~L.}\ \bibnamefont {Holloway}},\ }\bibfield  {title} {\bibinfo {title} {Rydberg atom-based sensors for radio-frequency electric field metrology, sensing, and communications},\ }\href {https://doi.org/https://doi.org/10.1016/j.measen.2021.100273} {\bibfield  {journal} {\bibinfo  {journal} {Measurement: Sensors}\ }\textbf {\bibinfo {volume} {18}},\ \bibinfo {pages} {100273} (\bibinfo {year} {2021}{\natexlab{a}})}\BibitemShut {NoStop}%
\bibitem [{\citenamefont {Simons}\ \emph {et~al.}(2021{\natexlab{b}})\citenamefont {Simons}, \citenamefont {Artusio-Glimpse}, \citenamefont {Holloway}, \citenamefont {Imhof}, \citenamefont {Jefferts}, \citenamefont {Wyllie}, \citenamefont {Sawyer},\ and\ \citenamefont {Walker}}]{Simons2021_2}%
  \BibitemOpen
  \bibfield  {author} {\bibinfo {author} {\bibfnamefont {M.~T.}\ \bibnamefont {Simons}}, \bibinfo {author} {\bibfnamefont {A.~B.}\ \bibnamefont {Artusio-Glimpse}}, \bibinfo {author} {\bibfnamefont {C.~L.}\ \bibnamefont {Holloway}}, \bibinfo {author} {\bibfnamefont {E.}~\bibnamefont {Imhof}}, \bibinfo {author} {\bibfnamefont {S.~R.}\ \bibnamefont {Jefferts}}, \bibinfo {author} {\bibfnamefont {R.}~\bibnamefont {Wyllie}}, \bibinfo {author} {\bibfnamefont {B.~C.}\ \bibnamefont {Sawyer}},\ and\ \bibinfo {author} {\bibfnamefont {T.~G.}\ \bibnamefont {Walker}},\ }\bibfield  {title} {\bibinfo {title} {Continuous radio-frequency electric-field detection through adjacent rydberg resonance tuning},\ }\href {https://doi.org/10.1103/PhysRevA.104.032824} {\bibfield  {journal} {\bibinfo  {journal} {Phys. Rev. A}\ }\textbf {\bibinfo {volume} {104}},\ \bibinfo {pages} {032824} (\bibinfo {year} {2021}{\natexlab{b}})}\BibitemShut {NoStop}%
\bibitem [{\citenamefont {Yuan}\ \emph {et~al.}(2023)\citenamefont {Yuan}, \citenamefont {Yang}, \citenamefont {Jing}, \citenamefont {Zhang}, \citenamefont {Jiao}, \citenamefont {Li}, \citenamefont {Zhang}, \citenamefont {Xiao},\ and\ \citenamefont {Jia}}]{Yuan2023}%
  \BibitemOpen
  \bibfield  {author} {\bibinfo {author} {\bibfnamefont {J.}~\bibnamefont {Yuan}}, \bibinfo {author} {\bibfnamefont {W.}~\bibnamefont {Yang}}, \bibinfo {author} {\bibfnamefont {M.}~\bibnamefont {Jing}}, \bibinfo {author} {\bibfnamefont {H.}~\bibnamefont {Zhang}}, \bibinfo {author} {\bibfnamefont {Y.}~\bibnamefont {Jiao}}, \bibinfo {author} {\bibfnamefont {W.}~\bibnamefont {Li}}, \bibinfo {author} {\bibfnamefont {L.}~\bibnamefont {Zhang}}, \bibinfo {author} {\bibfnamefont {L.}~\bibnamefont {Xiao}},\ and\ \bibinfo {author} {\bibfnamefont {S.}~\bibnamefont {Jia}},\ }\bibfield  {title} {\bibinfo {title} {Quantum sensing of microwave electric fields based on rydberg atoms},\ }\href {https://doi.org/10.1088/1361-6633/acf22f} {\bibfield  {journal} {\bibinfo  {journal} {Reports on Progress in Physics}\ }\textbf {\bibinfo {volume} {86}},\ \bibinfo {pages} {106001} (\bibinfo {year} {2023})}\BibitemShut {NoStop}%
\bibitem [{\citenamefont {Ovsiannikov}\ \emph {et~al.}(2022)\citenamefont {Ovsiannikov}, \citenamefont {Palchikov},\ and\ \citenamefont {Glukhov}}]{Ovsiannikov2022}%
  \BibitemOpen
  \bibfield  {author} {\bibinfo {author} {\bibfnamefont {V.~D.}\ \bibnamefont {Ovsiannikov}}, \bibinfo {author} {\bibfnamefont {V.~G.}\ \bibnamefont {Palchikov}},\ and\ \bibinfo {author} {\bibfnamefont {I.~L.}\ \bibnamefont {Glukhov}},\ }\bibfield  {title} {\bibinfo {title} {Microwave field metrology based on rydberg states of alkali-metal atoms},\ }\bibfield  {journal} {\bibinfo  {journal} {Photonics}\ }\textbf {\bibinfo {volume} {9}},\ \href {https://doi.org/10.3390/photonics9090635} {10.3390/photonics9090635} (\bibinfo {year} {2022})\BibitemShut {NoStop}%
\bibitem [{\citenamefont {Fahey}\ and\ \citenamefont {Noel}(2011)}]{Fahey2011}%
  \BibitemOpen
  \bibfield  {author} {\bibinfo {author} {\bibfnamefont {D.~P.}\ \bibnamefont {Fahey}}\ and\ \bibinfo {author} {\bibfnamefont {M.~W.}\ \bibnamefont {Noel}},\ }\bibfield  {title} {\bibinfo {title} {Excitation of rydberg states in rubidium with near infrared diode lasers},\ }\href {https://doi.org/10.1364/OE.19.017002} {\bibfield  {journal} {\bibinfo  {journal} {Opt. Express}\ }\textbf {\bibinfo {volume} {19}},\ \bibinfo {pages} {17002} (\bibinfo {year} {2011})}\BibitemShut {NoStop}%
\bibitem [{\citenamefont {Gorniaczyk}\ \emph {et~al.}(2014)\citenamefont {Gorniaczyk}, \citenamefont {Tresp}, \citenamefont {Schmidt}, \citenamefont {Fedder},\ and\ \citenamefont {Hofferberth}}]{Gorniaczyk2014}%
  \BibitemOpen
  \bibfield  {author} {\bibinfo {author} {\bibfnamefont {H.}~\bibnamefont {Gorniaczyk}}, \bibinfo {author} {\bibfnamefont {C.}~\bibnamefont {Tresp}}, \bibinfo {author} {\bibfnamefont {J.}~\bibnamefont {Schmidt}}, \bibinfo {author} {\bibfnamefont {H.}~\bibnamefont {Fedder}},\ and\ \bibinfo {author} {\bibfnamefont {S.}~\bibnamefont {Hofferberth}},\ }\bibfield  {title} {\bibinfo {title} {Single-photon transistor mediated by interstate rydberg interactions},\ }\href {https://doi.org/10.1103/PhysRevLett.113.053601} {\bibfield  {journal} {\bibinfo  {journal} {Phys. Rev. Lett.}\ }\textbf {\bibinfo {volume} {113}},\ \bibinfo {pages} {053601} (\bibinfo {year} {2014})}\BibitemShut {NoStop}%
\bibitem [{\citenamefont {Ryabtsev}\ \emph {et~al.}(2016)\citenamefont {Ryabtsev}, \citenamefont {Beterov}, \citenamefont {Tretyakov}, \citenamefont {Entin},\ and\ \citenamefont {Yakshina}}]{Ryabtsev2016}%
  \BibitemOpen
  \bibfield  {author} {\bibinfo {author} {\bibfnamefont {I.~I.}\ \bibnamefont {Ryabtsev}}, \bibinfo {author} {\bibfnamefont {I.~I.}\ \bibnamefont {Beterov}}, \bibinfo {author} {\bibfnamefont {D.~B.}\ \bibnamefont {Tretyakov}}, \bibinfo {author} {\bibfnamefont {V.~M.}\ \bibnamefont {Entin}},\ and\ \bibinfo {author} {\bibfnamefont {E.~A.}\ \bibnamefont {Yakshina}},\ }\bibfield  {title} {\bibinfo {title} {Spectroscopy of cold rubidium rydberg atoms for applications in quantum information},\ }\href {https://doi.org/10.3367/UFNe.0186.201602k.0206} {\bibfield  {journal} {\bibinfo  {journal} {Physics-Uspekhi}\ }\textbf {\bibinfo {volume} {59}},\ \bibinfo {pages} {196} (\bibinfo {year} {2016})}\BibitemShut {NoStop}%
\bibitem [{\citenamefont {Saffman}(2016)}]{Saffman2016}%
  \BibitemOpen
  \bibfield  {author} {\bibinfo {author} {\bibfnamefont {M.}~\bibnamefont {Saffman}},\ }\bibfield  {title} {\bibinfo {title} {Quantum computing with atomic qubits and rydberg interactions: progress and challenges},\ }\href {https://doi.org/10.1088/0953-4075/49/20/202001} {\bibfield  {journal} {\bibinfo  {journal} {Journal of Physics B: Atomic, Molecular and Optical Physics}\ }\textbf {\bibinfo {volume} {49}},\ \bibinfo {pages} {202001} (\bibinfo {year} {2016})}\BibitemShut {NoStop}%
\bibitem [{\citenamefont {Levine}\ \emph {et~al.}(2018)\citenamefont {Levine}, \citenamefont {Keesling}, \citenamefont {Omran}, \citenamefont {Bernien}, \citenamefont {Schwartz}, \citenamefont {Zibrov}, \citenamefont {Endres}, \citenamefont {Greiner}, \citenamefont {Vuleti\ifmmode~\acute{c}\else \'{c}\fi{}},\ and\ \citenamefont {Lukin}}]{Levine2018}%
  \BibitemOpen
  \bibfield  {author} {\bibinfo {author} {\bibfnamefont {H.}~\bibnamefont {Levine}}, \bibinfo {author} {\bibfnamefont {A.}~\bibnamefont {Keesling}}, \bibinfo {author} {\bibfnamefont {A.}~\bibnamefont {Omran}}, \bibinfo {author} {\bibfnamefont {H.}~\bibnamefont {Bernien}}, \bibinfo {author} {\bibfnamefont {S.}~\bibnamefont {Schwartz}}, \bibinfo {author} {\bibfnamefont {A.~S.}\ \bibnamefont {Zibrov}}, \bibinfo {author} {\bibfnamefont {M.}~\bibnamefont {Endres}}, \bibinfo {author} {\bibfnamefont {M.}~\bibnamefont {Greiner}}, \bibinfo {author} {\bibfnamefont {V.}~\bibnamefont {Vuleti\ifmmode~\acute{c}\else \'{c}\fi{}}},\ and\ \bibinfo {author} {\bibfnamefont {M.~D.}\ \bibnamefont {Lukin}},\ }\bibfield  {title} {\bibinfo {title} {High-fidelity control and entanglement of rydberg-atom qubits},\ }\href {https://doi.org/10.1103/PhysRevLett.121.123603} {\bibfield  {journal} {\bibinfo  {journal} {Phys. Rev. Lett.}\ }\textbf {\bibinfo {volume} {121}},\ \bibinfo {pages} {123603} (\bibinfo {year} {2018})}\BibitemShut
  {NoStop}%
\bibitem [{\citenamefont {Duspayev}\ \emph {et~al.}(2021)\citenamefont {Duspayev}, \citenamefont {Han}, \citenamefont {Viray}, \citenamefont {Ma}, \citenamefont {Zhao},\ and\ \citenamefont {Raithel}}]{Duspayev2021}%
  \BibitemOpen
  \bibfield  {author} {\bibinfo {author} {\bibfnamefont {A.}~\bibnamefont {Duspayev}}, \bibinfo {author} {\bibfnamefont {X.}~\bibnamefont {Han}}, \bibinfo {author} {\bibfnamefont {M.~A.}\ \bibnamefont {Viray}}, \bibinfo {author} {\bibfnamefont {L.}~\bibnamefont {Ma}}, \bibinfo {author} {\bibfnamefont {J.}~\bibnamefont {Zhao}},\ and\ \bibinfo {author} {\bibfnamefont {G.}~\bibnamefont {Raithel}},\ }\bibfield  {title} {\bibinfo {title} {Long-range rydberg-atom--ion molecules of rb and cs},\ }\href {https://doi.org/10.1103/PhysRevResearch.3.023114} {\bibfield  {journal} {\bibinfo  {journal} {Phys. Rev. Res.}\ }\textbf {\bibinfo {volume} {3}},\ \bibinfo {pages} {023114} (\bibinfo {year} {2021})}\BibitemShut {NoStop}%
\bibitem [{\citenamefont {Delteil}\ \emph {et~al.}(2019)\citenamefont {Delteil}, \citenamefont {Fink}, \citenamefont {Schade}, \citenamefont {H{\"o}fling}, \citenamefont {Schneider},\ and\ \citenamefont {{\.{I}}mamo{\u{g}}lu}}]{Delteil2019}%
  \BibitemOpen
  \bibfield  {author} {\bibinfo {author} {\bibfnamefont {A.}~\bibnamefont {Delteil}}, \bibinfo {author} {\bibfnamefont {T.}~\bibnamefont {Fink}}, \bibinfo {author} {\bibfnamefont {A.}~\bibnamefont {Schade}}, \bibinfo {author} {\bibfnamefont {S.}~\bibnamefont {H{\"o}fling}}, \bibinfo {author} {\bibfnamefont {C.}~\bibnamefont {Schneider}},\ and\ \bibinfo {author} {\bibfnamefont {A.}~\bibnamefont {{\.{I}}mamo{\u{g}}lu}},\ }\bibfield  {title} {\bibinfo {title} {Towards polariton blockade of confined exciton--polaritons},\ }\href {https://doi.org/10.1038/s41563-019-0282-y} {\bibfield  {journal} {\bibinfo  {journal} {Nature Materials}\ }\textbf {\bibinfo {volume} {18}},\ \bibinfo {pages} {219} (\bibinfo {year} {2019})}\BibitemShut {NoStop}%
\bibitem [{\citenamefont {Deligiannis}\ \emph {et~al.}(2022)\citenamefont {Deligiannis}, \citenamefont {Fontaine}, \citenamefont {Squizzato}, \citenamefont {Richard}, \citenamefont {Ravets}, \citenamefont {Bloch}, \citenamefont {Minguzzi},\ and\ \citenamefont {Canet}}]{Deligiannis2022}%
  \BibitemOpen
  \bibfield  {author} {\bibinfo {author} {\bibfnamefont {K.}~\bibnamefont {Deligiannis}}, \bibinfo {author} {\bibfnamefont {Q.}~\bibnamefont {Fontaine}}, \bibinfo {author} {\bibfnamefont {D.}~\bibnamefont {Squizzato}}, \bibinfo {author} {\bibfnamefont {M.}~\bibnamefont {Richard}}, \bibinfo {author} {\bibfnamefont {S.}~\bibnamefont {Ravets}}, \bibinfo {author} {\bibfnamefont {J.}~\bibnamefont {Bloch}}, \bibinfo {author} {\bibfnamefont {A.}~\bibnamefont {Minguzzi}},\ and\ \bibinfo {author} {\bibfnamefont {L.}~\bibnamefont {Canet}},\ }\bibfield  {title} {\bibinfo {title} {Kardar-parisi-zhang universality in discrete two-dimensional driven-dissipative exciton polariton condensates},\ }\href {https://doi.org/10.1103/PhysRevResearch.4.043207} {\bibfield  {journal} {\bibinfo  {journal} {Phys. Rev. Res.}\ }\textbf {\bibinfo {volume} {4}},\ \bibinfo {pages} {043207} (\bibinfo {year} {2022})}\BibitemShut {NoStop}%
\bibitem [{\citenamefont {Bloch}\ \emph {et~al.}(2022)\citenamefont {Bloch}, \citenamefont {Carusotto},\ and\ \citenamefont {Wouters}}]{Bloch2022}%
  \BibitemOpen
  \bibfield  {author} {\bibinfo {author} {\bibfnamefont {J.}~\bibnamefont {Bloch}}, \bibinfo {author} {\bibfnamefont {I.}~\bibnamefont {Carusotto}},\ and\ \bibinfo {author} {\bibfnamefont {M.}~\bibnamefont {Wouters}},\ }\bibfield  {title} {\bibinfo {title} {Non-equilibrium bose--einstein condensation in photonic systems},\ }\href {https://doi.org/10.1038/s42254-022-00464-0} {\bibfield  {journal} {\bibinfo  {journal} {Nature Reviews Physics}\ }\textbf {\bibinfo {volume} {4}},\ \bibinfo {pages} {470} (\bibinfo {year} {2022})}\BibitemShut {NoStop}%
\bibitem [{\citenamefont {Xiao}\ \emph {et~al.}(2017)\citenamefont {Xiao}, \citenamefont {Zhao}, \citenamefont {Wang},\ and\ \citenamefont {Zhang}}]{Xiao2017}%
  \BibitemOpen
  \bibfield  {author} {\bibinfo {author} {\bibfnamefont {J.}~\bibnamefont {Xiao}}, \bibinfo {author} {\bibfnamefont {M.}~\bibnamefont {Zhao}}, \bibinfo {author} {\bibfnamefont {Y.}~\bibnamefont {Wang}},\ and\ \bibinfo {author} {\bibfnamefont {X.}~\bibnamefont {Zhang}},\ }\bibfield  {title} {\bibinfo {title} {Excitons in atomically thin 2d semiconductors and their applications},\ }\href {https://doi.org/doi:10.1515/nanoph-2016-0160} {\bibfield  {journal} {\bibinfo  {journal} {Nanophotonics}\ }\textbf {\bibinfo {volume} {6}},\ \bibinfo {pages} {1309} (\bibinfo {year} {2017})}\BibitemShut {NoStop}%
\bibitem [{\citenamefont {Mueller}\ and\ \citenamefont {Malic}(2018)}]{Mueller2018}%
  \BibitemOpen
  \bibfield  {author} {\bibinfo {author} {\bibfnamefont {T.}~\bibnamefont {Mueller}}\ and\ \bibinfo {author} {\bibfnamefont {E.}~\bibnamefont {Malic}},\ }\bibfield  {title} {\bibinfo {title} {Exciton physics and device application of two-dimensional transition metal dichalcogenide semiconductors},\ }\href {https://doi.org/10.1038/s41699-018-0074-2} {\bibfield  {journal} {\bibinfo  {journal} {npj 2D Materials and Applications}\ }\textbf {\bibinfo {volume} {2}},\ \bibinfo {pages} {29} (\bibinfo {year} {2018})}\BibitemShut {NoStop}%
\bibitem [{\citenamefont {Riis-Jensen}\ \emph {et~al.}(2020)\citenamefont {Riis-Jensen}, \citenamefont {Gjerding}, \citenamefont {Russo},\ and\ \citenamefont {Thygesen}}]{Riis2020}%
  \BibitemOpen
  \bibfield  {author} {\bibinfo {author} {\bibfnamefont {A.~C.}\ \bibnamefont {Riis-Jensen}}, \bibinfo {author} {\bibfnamefont {M.~N.}\ \bibnamefont {Gjerding}}, \bibinfo {author} {\bibfnamefont {S.}~\bibnamefont {Russo}},\ and\ \bibinfo {author} {\bibfnamefont {K.~S.}\ \bibnamefont {Thygesen}},\ }\bibfield  {title} {\bibinfo {title} {Anomalous exciton rydberg series in two-dimensional semiconductors on high-$\ensuremath{\kappa}$ dielectric substrates},\ }\href {https://doi.org/10.1103/PhysRevB.102.201402} {\bibfield  {journal} {\bibinfo  {journal} {Phys. Rev. B}\ }\textbf {\bibinfo {volume} {102}},\ \bibinfo {pages} {201402} (\bibinfo {year} {2020})}\BibitemShut {NoStop}%
\bibitem [{\citenamefont {Chaves}\ \emph {et~al.}(2023)\citenamefont {Chaves}, \citenamefont {Teles},\ and\ \citenamefont {Thomen}}]{Chaves2023}%
  \BibitemOpen
  \bibfield  {author} {\bibinfo {author} {\bibfnamefont {A.}~\bibnamefont {Chaves}}, \bibinfo {author} {\bibfnamefont {L.~K.}\ \bibnamefont {Teles}},\ and\ \bibinfo {author} {\bibfnamefont {D.~N.}\ \bibnamefont {Thomen}},\ }\bibfield  {title} {\bibinfo {title} {Chapter two - excitons in two-dimensional semiconductors and van der waals heterostructures},\ }in\ \href {https://doi.org/https://doi.org/10.1016/bs.ssp.2023.09.001} {\emph {\bibinfo {booktitle} {Solid State Physics}}},\ \bibinfo {series} {Solid State Physics}, Vol.~\bibinfo {volume} {74},\ \bibinfo {editor} {edited by\ \bibinfo {editor} {\bibfnamefont {R.}~\bibnamefont {Macedo}}\ and\ \bibinfo {editor} {\bibfnamefont {R.~L.}\ \bibnamefont {Stamps}}}\ (\bibinfo  {publisher} {Academic Press},\ \bibinfo {year} {2023})\ pp.\ \bibinfo {pages} {67--94}\BibitemShut {NoStop}%
\bibitem [{\citenamefont {Bao}\ \emph {et~al.}(2019)\citenamefont {Bao}, \citenamefont {Liu}, \citenamefont {Xue}, \citenamefont {Zheng}, \citenamefont {Tao}, \citenamefont {Wang}, \citenamefont {Xia}, \citenamefont {Zhao}, \citenamefont {Kim}, \citenamefont {Yang}, \citenamefont {Li}, \citenamefont {Wang}, \citenamefont {Wang}, \citenamefont {Wang}, \citenamefont {MacDonald},\ and\ \citenamefont {Zhang}}]{Bao2019}%
  \BibitemOpen
  \bibfield  {author} {\bibinfo {author} {\bibfnamefont {W.}~\bibnamefont {Bao}}, \bibinfo {author} {\bibfnamefont {X.}~\bibnamefont {Liu}}, \bibinfo {author} {\bibfnamefont {F.}~\bibnamefont {Xue}}, \bibinfo {author} {\bibfnamefont {F.}~\bibnamefont {Zheng}}, \bibinfo {author} {\bibfnamefont {R.}~\bibnamefont {Tao}}, \bibinfo {author} {\bibfnamefont {S.}~\bibnamefont {Wang}}, \bibinfo {author} {\bibfnamefont {Y.}~\bibnamefont {Xia}}, \bibinfo {author} {\bibfnamefont {M.}~\bibnamefont {Zhao}}, \bibinfo {author} {\bibfnamefont {J.}~\bibnamefont {Kim}}, \bibinfo {author} {\bibfnamefont {S.}~\bibnamefont {Yang}}, \bibinfo {author} {\bibfnamefont {Q.}~\bibnamefont {Li}}, \bibinfo {author} {\bibfnamefont {Y.}~\bibnamefont {Wang}}, \bibinfo {author} {\bibfnamefont {Y.}~\bibnamefont {Wang}}, \bibinfo {author} {\bibfnamefont {L.-W.}\ \bibnamefont {Wang}}, \bibinfo {author} {\bibfnamefont {A.~H.}\ \bibnamefont {MacDonald}},\ and\ \bibinfo {author} {\bibfnamefont {X.}~\bibnamefont {Zhang}},\ }\bibfield  {title}
  {\bibinfo {title} {Observation of rydberg exciton polaritons and their condensate in a perovskite cavity},\ }\href {https://doi.org/10.1073/pnas.1909948116} {\bibfield  {journal} {\bibinfo  {journal} {Proceedings of the National Academy of Sciences}\ }\textbf {\bibinfo {volume} {116}},\ \bibinfo {pages} {20274} (\bibinfo {year} {2019})}\BibitemShut {NoStop}%
\bibitem [{\citenamefont {Ahumada-Lazo}\ \emph {et~al.}(2021)\citenamefont {Ahumada-Lazo}, \citenamefont {Saran}, \citenamefont {Woolland}, \citenamefont {Jia}, \citenamefont {Kyriazi}, \citenamefont {Kanaras}, \citenamefont {Binks},\ and\ \citenamefont {Curry}}]{Ahumada-Lazo_2021}%
  \BibitemOpen
  \bibfield  {author} {\bibinfo {author} {\bibfnamefont {R.}~\bibnamefont {Ahumada-Lazo}}, \bibinfo {author} {\bibfnamefont {R.}~\bibnamefont {Saran}}, \bibinfo {author} {\bibfnamefont {O.}~\bibnamefont {Woolland}}, \bibinfo {author} {\bibfnamefont {Y.}~\bibnamefont {Jia}}, \bibinfo {author} {\bibfnamefont {M.-E.}\ \bibnamefont {Kyriazi}}, \bibinfo {author} {\bibfnamefont {A.~G.}\ \bibnamefont {Kanaras}}, \bibinfo {author} {\bibfnamefont {D.}~\bibnamefont {Binks}},\ and\ \bibinfo {author} {\bibfnamefont {R.~J.}\ \bibnamefont {Curry}},\ }\bibfield  {title} {\bibinfo {title} {Exciton effects in perovskite nanocrystals},\ }\href {https://doi.org/10.1088/2515-7647/abedd0} {\bibfield  {journal} {\bibinfo  {journal} {Journal of Physics: Photonics}\ }\textbf {\bibinfo {volume} {3}},\ \bibinfo {pages} {021002} (\bibinfo {year} {2021})}\BibitemShut {NoStop}%
\bibitem [{\citenamefont {Su}\ \emph {et~al.}(2021)\citenamefont {Su}, \citenamefont {Fieramosca}, \citenamefont {Zhang}, \citenamefont {Nguyen}, \citenamefont {Deleporte}, \citenamefont {Chen}, \citenamefont {Sanvitto}, \citenamefont {Liew},\ and\ \citenamefont {Xiong}}]{Su2021}%
  \BibitemOpen
  \bibfield  {author} {\bibinfo {author} {\bibfnamefont {R.}~\bibnamefont {Su}}, \bibinfo {author} {\bibfnamefont {A.}~\bibnamefont {Fieramosca}}, \bibinfo {author} {\bibfnamefont {Q.}~\bibnamefont {Zhang}}, \bibinfo {author} {\bibfnamefont {H.~S.}\ \bibnamefont {Nguyen}}, \bibinfo {author} {\bibfnamefont {E.}~\bibnamefont {Deleporte}}, \bibinfo {author} {\bibfnamefont {Z.}~\bibnamefont {Chen}}, \bibinfo {author} {\bibfnamefont {D.}~\bibnamefont {Sanvitto}}, \bibinfo {author} {\bibfnamefont {T.~C.~H.}\ \bibnamefont {Liew}},\ and\ \bibinfo {author} {\bibfnamefont {Q.}~\bibnamefont {Xiong}},\ }\bibfield  {title} {\bibinfo {title} {Perovskite semiconductors for room-temperature exciton-polaritonics},\ }\href {https://doi.org/10.1038/s41563-021-01035-x} {\bibfield  {journal} {\bibinfo  {journal} {Nature Materials}\ }\textbf {\bibinfo {volume} {20}},\ \bibinfo {pages} {1315} (\bibinfo {year} {2021})}\BibitemShut {NoStop}%
\bibitem [{\citenamefont {Versteegh}\ \emph {et~al.}(2021)\citenamefont {Versteegh}, \citenamefont {Steinhauer}, \citenamefont {Bajo}, \citenamefont {Lettner}, \citenamefont {Soro}, \citenamefont {Romanova}, \citenamefont {Gyger}, \citenamefont {Schweickert}, \citenamefont {Mysyrowicz},\ and\ \citenamefont {Zwiller}}]{Versteegh2021}%
  \BibitemOpen
  \bibfield  {author} {\bibinfo {author} {\bibfnamefont {M.~A.~M.}\ \bibnamefont {Versteegh}}, \bibinfo {author} {\bibfnamefont {S.}~\bibnamefont {Steinhauer}}, \bibinfo {author} {\bibfnamefont {J.}~\bibnamefont {Bajo}}, \bibinfo {author} {\bibfnamefont {T.}~\bibnamefont {Lettner}}, \bibinfo {author} {\bibfnamefont {A.}~\bibnamefont {Soro}}, \bibinfo {author} {\bibfnamefont {A.}~\bibnamefont {Romanova}}, \bibinfo {author} {\bibfnamefont {S.}~\bibnamefont {Gyger}}, \bibinfo {author} {\bibfnamefont {L.}~\bibnamefont {Schweickert}}, \bibinfo {author} {\bibfnamefont {A.}~\bibnamefont {Mysyrowicz}},\ and\ \bibinfo {author} {\bibfnamefont {V.}~\bibnamefont {Zwiller}},\ }\bibfield  {title} {\bibinfo {title} {Giant rydberg excitons in ${\mathrm{cu}}_{2}\mathrm{O}$ probed by photoluminescence excitation spectroscopy},\ }\href {https://doi.org/10.1103/PhysRevB.104.245206} {\bibfield  {journal} {\bibinfo  {journal} {Phys. Rev. B}\ }\textbf {\bibinfo {volume} {104}},\ \bibinfo {pages} {245206} (\bibinfo {year}
  {2021})}\BibitemShut {NoStop}%
\bibitem [{\citenamefont {Takahata}\ and\ \citenamefont {Naka}(2018)}]{Takahata2018}%
  \BibitemOpen
  \bibfield  {author} {\bibinfo {author} {\bibfnamefont {M.}~\bibnamefont {Takahata}}\ and\ \bibinfo {author} {\bibfnamefont {N.}~\bibnamefont {Naka}},\ }\bibfield  {title} {\bibinfo {title} {Photoluminescence properties of the entire excitonic series in ${\mathrm{cu}}_{2}\mathrm{O}$},\ }\href {https://doi.org/10.1103/PhysRevB.98.195205} {\bibfield  {journal} {\bibinfo  {journal} {Phys. Rev. B}\ }\textbf {\bibinfo {volume} {98}},\ \bibinfo {pages} {195205} (\bibinfo {year} {2018})}\BibitemShut {NoStop}%
\bibitem [{\citenamefont {Jang}(2015)}]{Jang2015}%
  \BibitemOpen
  \bibfield  {author} {\bibinfo {author} {\bibfnamefont {J.}~\bibnamefont {Jang}},\ }\bibfield  {title} {\bibinfo {title} {Lifetimes of excitons in cuprous oxide},\ }\href@noop {} {\bibfield  {journal} {\bibinfo  {journal} {Philosophical Magazine Series 1}\ } (\bibinfo {year} {2015})}\BibitemShut {NoStop}%
\bibitem [{\citenamefont {Aßmann}\ and\ \citenamefont {Bayer}(2020)}]{Asman2020}%
  \BibitemOpen
  \bibfield  {author} {\bibinfo {author} {\bibfnamefont {M.}~\bibnamefont {Aßmann}}\ and\ \bibinfo {author} {\bibfnamefont {M.}~\bibnamefont {Bayer}},\ }\bibfield  {title} {\bibinfo {title} {Semiconductor rydberg physics},\ }\href {https://doi.org/https://doi.org/10.1002/qute.201900134} {\bibfield  {journal} {\bibinfo  {journal} {Advanced Quantum Technologies}\ }\textbf {\bibinfo {volume} {3}},\ \bibinfo {pages} {1900134} (\bibinfo {year} {2020})}\BibitemShut {NoStop}%
\bibitem [{\citenamefont {Kazimierczuk}\ \emph {et~al.}(2014)\citenamefont {Kazimierczuk}, \citenamefont {Fr{\"o}hlich}, \citenamefont {Scheel}, \citenamefont {Stolz},\ and\ \citenamefont {Bayer}}]{Kazimierczuk2014}%
  \BibitemOpen
  \bibfield  {author} {\bibinfo {author} {\bibfnamefont {T.}~\bibnamefont {Kazimierczuk}}, \bibinfo {author} {\bibfnamefont {D.}~\bibnamefont {Fr{\"o}hlich}}, \bibinfo {author} {\bibfnamefont {S.}~\bibnamefont {Scheel}}, \bibinfo {author} {\bibfnamefont {H.}~\bibnamefont {Stolz}},\ and\ \bibinfo {author} {\bibfnamefont {M.}~\bibnamefont {Bayer}},\ }\bibfield  {title} {\bibinfo {title} {Giant rydberg excitons in the copper oxide cu2o},\ }\href {https://doi.org/10.1038/nature13832} {\bibfield  {journal} {\bibinfo  {journal} {Nature}\ }\textbf {\bibinfo {volume} {514}},\ \bibinfo {pages} {343} (\bibinfo {year} {2014})}\BibitemShut {NoStop}%
\bibitem [{\citenamefont {Kitamura}\ \emph {et~al.}(2017)\citenamefont {Kitamura}, \citenamefont {Takahata},\ and\ \citenamefont {Naka}}]{KITAMURA2017808}%
  \BibitemOpen
  \bibfield  {author} {\bibinfo {author} {\bibfnamefont {T.}~\bibnamefont {Kitamura}}, \bibinfo {author} {\bibfnamefont {M.}~\bibnamefont {Takahata}},\ and\ \bibinfo {author} {\bibfnamefont {N.}~\bibnamefont {Naka}},\ }\bibfield  {title} {\bibinfo {title} {Quantum number dependence of the photoluminescence broadening of excitonic rydberg states in cuprous oxide},\ }\href {https://doi.org/https://doi.org/10.1016/j.jlumin.2017.07.060} {\bibfield  {journal} {\bibinfo  {journal} {Journal of Luminescence}\ }\textbf {\bibinfo {volume} {192}},\ \bibinfo {pages} {808} (\bibinfo {year} {2017})}\BibitemShut {NoStop}%
\bibitem [{\citenamefont {Takahata}\ \emph {et~al.}(2018)\citenamefont {Takahata}, \citenamefont {Tanaka},\ and\ \citenamefont {Naka}}]{Takahata2018May}%
  \BibitemOpen
  \bibfield  {author} {\bibinfo {author} {\bibfnamefont {M.}~\bibnamefont {Takahata}}, \bibinfo {author} {\bibfnamefont {K.}~\bibnamefont {Tanaka}},\ and\ \bibinfo {author} {\bibfnamefont {N.}~\bibnamefont {Naka}},\ }\bibfield  {title} {\bibinfo {title} {Nonlocal optical response of weakly confined excitons in ${\mathrm{cu}}_{2}\mathrm{O}$ mesoscopic films},\ }\href {https://doi.org/10.1103/PhysRevB.97.205305} {\bibfield  {journal} {\bibinfo  {journal} {Phys. Rev. B}\ }\textbf {\bibinfo {volume} {97}},\ \bibinfo {pages} {205305} (\bibinfo {year} {2018})}\BibitemShut {NoStop}%
\bibitem [{\citenamefont {Naka}\ and\ \citenamefont {Nagasawa}(2001)}]{NAKA2001413}%
  \BibitemOpen
  \bibfield  {author} {\bibinfo {author} {\bibfnamefont {N.}~\bibnamefont {Naka}}\ and\ \bibinfo {author} {\bibfnamefont {N.}~\bibnamefont {Nagasawa}},\ }\bibfield  {title} {\bibinfo {title} {Dynamics of paraexcitons generated in a 3d confined potential well by two-photon resonance excitation in cu2o},\ }\href {https://doi.org/https://doi.org/10.1016/S0022-2313(01)00315-5} {\bibfield  {journal} {\bibinfo  {journal} {Journal of Luminescence}\ }\textbf {\bibinfo {volume} {94-95}},\ \bibinfo {pages} {413} (\bibinfo {year} {2001})},\ \bibinfo {note} {international Conference on Dynamical Processes in Excited States of Solids}\BibitemShut {NoStop}%
\bibitem [{\citenamefont {Paul}\ \emph {et~al.}(2024)\citenamefont {Paul}, \citenamefont {Rajendran}, \citenamefont {Ziemkiewicz}, \citenamefont {Volz},\ and\ \citenamefont {Ohadi}}]{Paul2024}%
  \BibitemOpen
  \bibfield  {author} {\bibinfo {author} {\bibfnamefont {A.~S.}\ \bibnamefont {Paul}}, \bibinfo {author} {\bibfnamefont {S.~K.}\ \bibnamefont {Rajendran}}, \bibinfo {author} {\bibfnamefont {D.}~\bibnamefont {Ziemkiewicz}}, \bibinfo {author} {\bibfnamefont {T.}~\bibnamefont {Volz}},\ and\ \bibinfo {author} {\bibfnamefont {H.}~\bibnamefont {Ohadi}},\ }\bibfield  {title} {\bibinfo {title} {Local tuning of rydberg exciton energies in nanofabricated cu2o pillars},\ }\href {https://doi.org/10.1038/s43246-024-00481-9} {\bibfield  {journal} {\bibinfo  {journal} {Communications Materials}\ }\textbf {\bibinfo {volume} {5}},\ \bibinfo {pages} {43} (\bibinfo {year} {2024})}\BibitemShut {NoStop}%
\bibitem [{\citenamefont {Koirala}\ \emph {et~al.}(2013)\citenamefont {Koirala}, \citenamefont {Naka},\ and\ \citenamefont {Tanaka}}]{Koirala2013}%
  \BibitemOpen
  \bibfield  {author} {\bibinfo {author} {\bibfnamefont {S.}~\bibnamefont {Koirala}}, \bibinfo {author} {\bibfnamefont {N.}~\bibnamefont {Naka}},\ and\ \bibinfo {author} {\bibfnamefont {K.}~\bibnamefont {Tanaka}},\ }\bibfield  {title} {\bibinfo {title} {Correlated lifetimes of free paraexcitons and excitons trapped at oxygen vacancies in cuprous oxide},\ }\href@noop {} {\bibfield  {journal} {\bibinfo  {journal} {Journal of Luminescence}\ }\textbf {\bibinfo {volume} {134}},\ \bibinfo {pages} {524} (\bibinfo {year} {2013})}\BibitemShut {NoStop}%
\bibitem [{\citenamefont {Steinhauer}\ \emph {et~al.}(2020)\citenamefont {Steinhauer}, \citenamefont {Versteegh}, \citenamefont {Gyger}, \citenamefont {Elshaari}, \citenamefont {Kunert}, \citenamefont {Mysyrowicz},\ and\ \citenamefont {Zwiller}}]{Steinhauer2020}%
  \BibitemOpen
  \bibfield  {author} {\bibinfo {author} {\bibfnamefont {S.}~\bibnamefont {Steinhauer}}, \bibinfo {author} {\bibfnamefont {M.~A.~M.}\ \bibnamefont {Versteegh}}, \bibinfo {author} {\bibfnamefont {S.}~\bibnamefont {Gyger}}, \bibinfo {author} {\bibfnamefont {A.~W.}\ \bibnamefont {Elshaari}}, \bibinfo {author} {\bibfnamefont {B.}~\bibnamefont {Kunert}}, \bibinfo {author} {\bibfnamefont {A.}~\bibnamefont {Mysyrowicz}},\ and\ \bibinfo {author} {\bibfnamefont {V.}~\bibnamefont {Zwiller}},\ }\bibfield  {title} {\bibinfo {title} {Rydberg excitons in cu2o microcrystals grown on a silicon platform},\ }\href {https://doi.org/10.1038/s43246-020-0013-6} {\bibfield  {journal} {\bibinfo  {journal} {Communications Materials}\ }\textbf {\bibinfo {volume} {1}},\ \bibinfo {pages} {11} (\bibinfo {year} {2020})}\BibitemShut {NoStop}%
\bibitem [{\citenamefont {DeLange}\ \emph {et~al.}(2023)\citenamefont {DeLange}, \citenamefont {Barua}, \citenamefont {Paul}, \citenamefont {Ohadi}, \citenamefont {Zwiller}, \citenamefont {Steinhauer},\ and\ \citenamefont {Alaeian}}]{DeLange2023}%
  \BibitemOpen
  \bibfield  {author} {\bibinfo {author} {\bibfnamefont {J.}~\bibnamefont {DeLange}}, \bibinfo {author} {\bibfnamefont {K.}~\bibnamefont {Barua}}, \bibinfo {author} {\bibfnamefont {A.~S.}\ \bibnamefont {Paul}}, \bibinfo {author} {\bibfnamefont {H.}~\bibnamefont {Ohadi}}, \bibinfo {author} {\bibfnamefont {V.}~\bibnamefont {Zwiller}}, \bibinfo {author} {\bibfnamefont {S.}~\bibnamefont {Steinhauer}},\ and\ \bibinfo {author} {\bibfnamefont {H.}~\bibnamefont {Alaeian}},\ }\bibfield  {title} {\bibinfo {title} {Highly-excited rydberg excitons in synthetic thin-film cuprous oxide},\ }\href {https://doi.org/10.1038/s41598-023-41465-y} {\bibfield  {journal} {\bibinfo  {journal} {Scientific Reports}\ }\textbf {\bibinfo {volume} {13}},\ \bibinfo {pages} {16881} (\bibinfo {year} {2023})}\BibitemShut {NoStop}%
\bibitem [{\citenamefont {Belov}\ \emph {et~al.}(2024)\citenamefont {Belov}, \citenamefont {Morawetz}, \citenamefont {Krüger}, \citenamefont {Scheuler}, \citenamefont {Rommel}, \citenamefont {Main}, \citenamefont {Giessen},\ and\ \citenamefont {Scheel}}]{Belov2024}%
  \BibitemOpen
  \bibfield  {author} {\bibinfo {author} {\bibfnamefont {P.~A.}\ \bibnamefont {Belov}}, \bibinfo {author} {\bibfnamefont {F.}~\bibnamefont {Morawetz}}, \bibinfo {author} {\bibfnamefont {S.~O.}\ \bibnamefont {Krüger}}, \bibinfo {author} {\bibfnamefont {N.}~\bibnamefont {Scheuler}}, \bibinfo {author} {\bibfnamefont {P.}~\bibnamefont {Rommel}}, \bibinfo {author} {\bibfnamefont {J.}~\bibnamefont {Main}}, \bibinfo {author} {\bibfnamefont {H.}~\bibnamefont {Giessen}},\ and\ \bibinfo {author} {\bibfnamefont {S.}~\bibnamefont {Scheel}},\ }\href@noop {} {\bibinfo {title} {Energy states of rydberg excitons in finite crystals: from weak to strong confinement}} (\bibinfo {year} {2024}),\ \Eprint {https://arxiv.org/abs/2310.19746} {arXiv:2310.19746 [cond-mat.mes-hall]} \BibitemShut {NoStop}%
\bibitem [{\citenamefont {Moon}\ \emph {et~al.}(2020)\citenamefont {Moon}, \citenamefont {Bersin}, \citenamefont {Chakraborty}, \citenamefont {Lu}, \citenamefont {Grosso}, \citenamefont {Kong},\ and\ \citenamefont {Englund}}]{Moon2020}%
  \BibitemOpen
  \bibfield  {author} {\bibinfo {author} {\bibfnamefont {H.}~\bibnamefont {Moon}}, \bibinfo {author} {\bibfnamefont {E.}~\bibnamefont {Bersin}}, \bibinfo {author} {\bibfnamefont {C.}~\bibnamefont {Chakraborty}}, \bibinfo {author} {\bibfnamefont {A.-Y.}\ \bibnamefont {Lu}}, \bibinfo {author} {\bibfnamefont {G.}~\bibnamefont {Grosso}}, \bibinfo {author} {\bibfnamefont {J.}~\bibnamefont {Kong}},\ and\ \bibinfo {author} {\bibfnamefont {D.}~\bibnamefont {Englund}},\ }\bibfield  {title} {\bibinfo {title} {Strain-correlated localized exciton energy in atomically thin semiconductors},\ }\href {https://doi.org/10.1021/acsphotonics.0c00626} {\bibfield  {journal} {\bibinfo  {journal} {ACS Photonics}\ }\textbf {\bibinfo {volume} {7}},\ \bibinfo {pages} {1135} (\bibinfo {year} {2020})}\BibitemShut {NoStop}%
\bibitem [{\citenamefont {Thureja}\ \emph {et~al.}(2022)\citenamefont {Thureja}, \citenamefont {Imamoglu}, \citenamefont {Smole{\'{n}}ski}, \citenamefont {Amelio}, \citenamefont {Popert}, \citenamefont {Chervy}, \citenamefont {Lu}, \citenamefont {Liu}, \citenamefont {Barmak}, \citenamefont {Watanabe}, \citenamefont {Taniguchi}, \citenamefont {Norris}, \citenamefont {Kroner},\ and\ \citenamefont {Murthy}}]{Thureja2022}%
  \BibitemOpen
  \bibfield  {author} {\bibinfo {author} {\bibfnamefont {D.}~\bibnamefont {Thureja}}, \bibinfo {author} {\bibfnamefont {A.}~\bibnamefont {Imamoglu}}, \bibinfo {author} {\bibfnamefont {T.}~\bibnamefont {Smole{\'{n}}ski}}, \bibinfo {author} {\bibfnamefont {I.}~\bibnamefont {Amelio}}, \bibinfo {author} {\bibfnamefont {A.}~\bibnamefont {Popert}}, \bibinfo {author} {\bibfnamefont {T.}~\bibnamefont {Chervy}}, \bibinfo {author} {\bibfnamefont {X.}~\bibnamefont {Lu}}, \bibinfo {author} {\bibfnamefont {S.}~\bibnamefont {Liu}}, \bibinfo {author} {\bibfnamefont {K.}~\bibnamefont {Barmak}}, \bibinfo {author} {\bibfnamefont {K.}~\bibnamefont {Watanabe}}, \bibinfo {author} {\bibfnamefont {T.}~\bibnamefont {Taniguchi}}, \bibinfo {author} {\bibfnamefont {D.~J.}\ \bibnamefont {Norris}}, \bibinfo {author} {\bibfnamefont {M.}~\bibnamefont {Kroner}},\ and\ \bibinfo {author} {\bibfnamefont {P.~A.}\ \bibnamefont {Murthy}},\ }\bibfield  {title} {\bibinfo {title} {Electrically tunable quantum confinement of neutral excitons},\
  }\href {https://doi.org/10.1038/s41586-022-04634-z} {\bibfield  {journal} {\bibinfo  {journal} {Nature}\ }\textbf {\bibinfo {volume} {606}},\ \bibinfo {pages} {298} (\bibinfo {year} {2022})}\BibitemShut {NoStop}%
\bibitem [{\citenamefont {Bajoni}\ \emph {et~al.}(2008)\citenamefont {Bajoni}, \citenamefont {Wertz}, \citenamefont {Senellart}, \citenamefont {Miard}, \citenamefont {Semenova}, \citenamefont {Lemaître}, \citenamefont {Sagnes}, \citenamefont {Bouchoule},\ and\ \citenamefont {Bloch}}]{Bajoni2008}%
  \BibitemOpen
  \bibfield  {author} {\bibinfo {author} {\bibfnamefont {D.}~\bibnamefont {Bajoni}}, \bibinfo {author} {\bibfnamefont {E.}~\bibnamefont {Wertz}}, \bibinfo {author} {\bibfnamefont {P.}~\bibnamefont {Senellart}}, \bibinfo {author} {\bibfnamefont {A.}~\bibnamefont {Miard}}, \bibinfo {author} {\bibfnamefont {E.}~\bibnamefont {Semenova}}, \bibinfo {author} {\bibfnamefont {A.}~\bibnamefont {Lemaître}}, \bibinfo {author} {\bibfnamefont {I.}~\bibnamefont {Sagnes}}, \bibinfo {author} {\bibfnamefont {S.}~\bibnamefont {Bouchoule}},\ and\ \bibinfo {author} {\bibfnamefont {J.}~\bibnamefont {Bloch}},\ }\bibfield  {title} {\bibinfo {title} {Excitonic polaritons in semiconductor micropillars},\ }\href {https://doi.org/10.12693/APhysPolA.114.933} {\bibfield  {journal} {\bibinfo  {journal} {Acta Physica Polonica A - ACTA PHYS POL A}\ }\textbf {\bibinfo {volume} {114}} (\bibinfo {year} {2008})}\BibitemShut {NoStop}%
\bibitem [{\citenamefont {Kuznetsov}\ \emph {et~al.}(2018)\citenamefont {Kuznetsov}, \citenamefont {Helgers}, \citenamefont {Biermann},\ and\ \citenamefont {Santos}}]{Kuznetsov2018}%
  \BibitemOpen
  \bibfield  {author} {\bibinfo {author} {\bibfnamefont {A.~S.}\ \bibnamefont {Kuznetsov}}, \bibinfo {author} {\bibfnamefont {P.~L.~J.}\ \bibnamefont {Helgers}}, \bibinfo {author} {\bibfnamefont {K.}~\bibnamefont {Biermann}},\ and\ \bibinfo {author} {\bibfnamefont {P.~V.}\ \bibnamefont {Santos}},\ }\bibfield  {title} {\bibinfo {title} {Quantum confinement of exciton-polaritons in a structured (al,ga)as microcavity},\ }\href {https://doi.org/10.1103/PhysRevB.97.195309} {\bibfield  {journal} {\bibinfo  {journal} {Phys. Rev. B}\ }\textbf {\bibinfo {volume} {97}},\ \bibinfo {pages} {195309} (\bibinfo {year} {2018})}\BibitemShut {NoStop}%
\bibitem [{\citenamefont {Orfanakis}\ \emph {et~al.}(2022)\citenamefont {Orfanakis}, \citenamefont {Rajendran}, \citenamefont {Walther}, \citenamefont {Volz}, \citenamefont {Pohl},\ and\ \citenamefont {Ohadi}}]{Orfanakis2022}%
  \BibitemOpen
  \bibfield  {author} {\bibinfo {author} {\bibfnamefont {K.}~\bibnamefont {Orfanakis}}, \bibinfo {author} {\bibfnamefont {S.~K.}\ \bibnamefont {Rajendran}}, \bibinfo {author} {\bibfnamefont {V.}~\bibnamefont {Walther}}, \bibinfo {author} {\bibfnamefont {T.}~\bibnamefont {Volz}}, \bibinfo {author} {\bibfnamefont {T.}~\bibnamefont {Pohl}},\ and\ \bibinfo {author} {\bibfnamefont {H.}~\bibnamefont {Ohadi}},\ }\bibfield  {title} {\bibinfo {title} {Rydberg exciton--polaritons in a cu2o microcavity},\ }\href {https://doi.org/10.1038/s41563-022-01230-4} {\bibfield  {journal} {\bibinfo  {journal} {Nature Materials}\ }\textbf {\bibinfo {volume} {21}},\ \bibinfo {pages} {767} (\bibinfo {year} {2022})}\BibitemShut {NoStop}%
\bibitem [{\citenamefont {Toyozawa}(1962)}]{Toyozawa1962}%
  \BibitemOpen
  \bibfield  {author} {\bibinfo {author} {\bibfnamefont {Y.}~\bibnamefont {Toyozawa}},\ }\bibfield  {title} {\bibinfo {title} {{Further Contribution to the Theory of the Line-Shape of the Exciton Absorption Band: }},\ }\href {https://doi.org/10.1143/PTP.27.89} {\bibfield  {journal} {\bibinfo  {journal} {Progress of Theoretical Physics}\ }\textbf {\bibinfo {volume} {27}},\ \bibinfo {pages} {89} (\bibinfo {year} {1962})}\BibitemShut {NoStop}%
\bibitem [{\citenamefont {Schweiner}\ \emph {et~al.}(2017)\citenamefont {Schweiner}, \citenamefont {Main}, \citenamefont {Wunner},\ and\ \citenamefont {Uihlein}}]{Schweiner2017}%
  \BibitemOpen
  \bibfield  {author} {\bibinfo {author} {\bibfnamefont {F.}~\bibnamefont {Schweiner}}, \bibinfo {author} {\bibfnamefont {J.}~\bibnamefont {Main}}, \bibinfo {author} {\bibfnamefont {G.}~\bibnamefont {Wunner}},\ and\ \bibinfo {author} {\bibfnamefont {C.}~\bibnamefont {Uihlein}},\ }\bibfield  {title} {\bibinfo {title} {Even exciton series in ${\mathrm{cu}}_{2}\mathrm{O}$},\ }\href {https://doi.org/10.1103/PhysRevB.95.195201} {\bibfield  {journal} {\bibinfo  {journal} {Phys. Rev. B}\ }\textbf {\bibinfo {volume} {95}},\ \bibinfo {pages} {195201} (\bibinfo {year} {2017})}\BibitemShut {NoStop}%
\bibitem [{\citenamefont {Toyozawa}(1964)}]{Toyozawa1964}%
  \BibitemOpen
  \bibfield  {author} {\bibinfo {author} {\bibfnamefont {Y.}~\bibnamefont {Toyozawa}},\ }\bibfield  {title} {\bibinfo {title} {Interband effect of lattice vibrations in the exciton absorption spectra},\ }\href {https://doi.org/https://doi.org/10.1016/0022-3697(64)90162-3} {\bibfield  {journal} {\bibinfo  {journal} {Journal of Physics and Chemistry of Solids}\ }\textbf {\bibinfo {volume} {25}},\ \bibinfo {pages} {59} (\bibinfo {year} {1964})}\BibitemShut {NoStop}%
\bibitem [{\citenamefont {Kr\"uger}\ \emph {et~al.}(2020)\citenamefont {Kr\"uger}, \citenamefont {Stolz},\ and\ \citenamefont {Scheel}}]{Kruger}%
  \BibitemOpen
  \bibfield  {author} {\bibinfo {author} {\bibfnamefont {S.~O.}\ \bibnamefont {Kr\"uger}}, \bibinfo {author} {\bibfnamefont {H.}~\bibnamefont {Stolz}},\ and\ \bibinfo {author} {\bibfnamefont {S.}~\bibnamefont {Scheel}},\ }\bibfield  {title} {\bibinfo {title} {Interaction of charged impurities and rydberg excitons in cuprous oxide},\ }\href {https://doi.org/10.1103/PhysRevB.101.235204} {\bibfield  {journal} {\bibinfo  {journal} {Phys. Rev. B}\ }\textbf {\bibinfo {volume} {101}},\ \bibinfo {pages} {235204} (\bibinfo {year} {2020})}\BibitemShut {NoStop}%
\bibitem [{\citenamefont {Sch\"one}\ \emph {et~al.}(2016)\citenamefont {Sch\"one}, \citenamefont {Kr\"uger}, \citenamefont {Gr\"unwald}, \citenamefont {Stolz}, \citenamefont {Scheel}, \citenamefont {A\ss{}mann}, \citenamefont {Heck\"otter}, \citenamefont {Thewes}, \citenamefont {Fr\"ohlich},\ and\ \citenamefont {Bayer}}]{Schone2016}%
  \BibitemOpen
  \bibfield  {author} {\bibinfo {author} {\bibfnamefont {F.}~\bibnamefont {Sch\"one}}, \bibinfo {author} {\bibfnamefont {S.-O.}\ \bibnamefont {Kr\"uger}}, \bibinfo {author} {\bibfnamefont {P.}~\bibnamefont {Gr\"unwald}}, \bibinfo {author} {\bibfnamefont {H.}~\bibnamefont {Stolz}}, \bibinfo {author} {\bibfnamefont {S.}~\bibnamefont {Scheel}}, \bibinfo {author} {\bibfnamefont {M.}~\bibnamefont {A\ss{}mann}}, \bibinfo {author} {\bibfnamefont {J.}~\bibnamefont {Heck\"otter}}, \bibinfo {author} {\bibfnamefont {J.}~\bibnamefont {Thewes}}, \bibinfo {author} {\bibfnamefont {D.}~\bibnamefont {Fr\"ohlich}},\ and\ \bibinfo {author} {\bibfnamefont {M.}~\bibnamefont {Bayer}},\ }\bibfield  {title} {\bibinfo {title} {Deviations of the exciton level spectrum in ${\mathrm{cu}}_{2}\mathrm{O}$ from the hydrogen series},\ }\href {https://doi.org/10.1103/PhysRevB.93.075203} {\bibfield  {journal} {\bibinfo  {journal} {Phys. Rev. B}\ }\textbf {\bibinfo {volume} {93}},\ \bibinfo {pages} {075203} (\bibinfo {year} {2016})}\BibitemShut
  {NoStop}%
\bibitem [{\citenamefont {Taylor}\ \emph {et~al.}(2022)\citenamefont {Taylor}, \citenamefont {Goswami}, \citenamefont {Walther}, \citenamefont {Spanner}, \citenamefont {Simon},\ and\ \citenamefont {Heshami}}]{Taylor_2022}%
  \BibitemOpen
  \bibfield  {author} {\bibinfo {author} {\bibfnamefont {J.}~\bibnamefont {Taylor}}, \bibinfo {author} {\bibfnamefont {S.}~\bibnamefont {Goswami}}, \bibinfo {author} {\bibfnamefont {V.}~\bibnamefont {Walther}}, \bibinfo {author} {\bibfnamefont {M.}~\bibnamefont {Spanner}}, \bibinfo {author} {\bibfnamefont {C.}~\bibnamefont {Simon}},\ and\ \bibinfo {author} {\bibfnamefont {K.}~\bibnamefont {Heshami}},\ }\bibfield  {title} {\bibinfo {title} {Simulation of many-body dynamics using rydberg excitons},\ }\href {https://doi.org/10.1088/2058-9565/ac70f4} {\bibfield  {journal} {\bibinfo  {journal} {Quantum Science and Technology}\ }\textbf {\bibinfo {volume} {7}},\ \bibinfo {pages} {035016} (\bibinfo {year} {2022})}\BibitemShut {NoStop}%
\bibitem [{\citenamefont {Neubauer}\ \emph {et~al.}(2022)\citenamefont {Neubauer}, \citenamefont {Heck\"otter}, \citenamefont {Ubl}, \citenamefont {Hentschel}, \citenamefont {Panda}, \citenamefont {A\ss{}mann}, \citenamefont {Bayer},\ and\ \citenamefont {Giessen}}]{Neubauer2022}%
  \BibitemOpen
  \bibfield  {author} {\bibinfo {author} {\bibfnamefont {A.}~\bibnamefont {Neubauer}}, \bibinfo {author} {\bibfnamefont {J.}~\bibnamefont {Heck\"otter}}, \bibinfo {author} {\bibfnamefont {M.}~\bibnamefont {Ubl}}, \bibinfo {author} {\bibfnamefont {M.}~\bibnamefont {Hentschel}}, \bibinfo {author} {\bibfnamefont {B.}~\bibnamefont {Panda}}, \bibinfo {author} {\bibfnamefont {M.}~\bibnamefont {A\ss{}mann}}, \bibinfo {author} {\bibfnamefont {M.}~\bibnamefont {Bayer}},\ and\ \bibinfo {author} {\bibfnamefont {H.}~\bibnamefont {Giessen}},\ }\bibfield  {title} {\bibinfo {title} {Spectroscopy of nanoantenna-covered ${\mathrm{cu}}_{2}\mathrm{O}$: Towards enhancing quadrupole transitions in rydberg excitons},\ }\href {https://doi.org/10.1103/PhysRevB.106.165305} {\bibfield  {journal} {\bibinfo  {journal} {Phys. Rev. B}\ }\textbf {\bibinfo {volume} {106}},\ \bibinfo {pages} {165305} (\bibinfo {year} {2022})}\BibitemShut {NoStop}%
\bibitem [{\citenamefont {Ziemkiewicz}\ and\ \citenamefont {Zieli\ifmmode \acute{n}\else \'{n}\fi{}ska-Raczy\ifmmode~\acute{n}\else \'{n}\fi{}ska}(2022)}]{Ziemkiewicz2022}%
  \BibitemOpen
  \bibfield  {author} {\bibinfo {author} {\bibfnamefont {D.}~\bibnamefont {Ziemkiewicz}}\ and\ \bibinfo {author} {\bibfnamefont {S.}~\bibnamefont {Zieli\ifmmode \acute{n}\else \'{n}\fi{}ska-Raczy\ifmmode~\acute{n}\else \'{n}\fi{}ska}},\ }\bibfield  {title} {\bibinfo {title} {Copper plasmonics with excitons},\ }\href {https://doi.org/10.1103/PhysRevB.106.205404} {\bibfield  {journal} {\bibinfo  {journal} {Phys. Rev. B}\ }\textbf {\bibinfo {volume} {106}},\ \bibinfo {pages} {205404} (\bibinfo {year} {2022})}\BibitemShut {NoStop}%
\end{thebibliography}%

\newpage

\onecolumngrid

\renewcommand{\appendix}{\par
  \setcounter{section}{0}
  \setcounter{subsection}{0}
  \setcounter{subsubsection}{0}
  \gdef\thesection{\Alph{section}}
  \gdef\thesubsection{\Alph{section}.\arabic{subsection}}
  \gdef\thesubsubsection{\Alph{section}.\arabic{subsection}.\roman{subsubsection}}
}

\appendix

\clearpage
% \maketitle
% \vspace{-20pt}
\begin{center} 
    \large {\textbf {Bottom-up Fabrication of 2D Rydberg Exciton Arrays in Cuprous Oxide\\}}
    \vspace{0.2in}
    \uppercase{\textbf {Supplemental Material}}
    \vspace{-0.1in}
\end{center}

\section{Actual Side Lengths of Square $\text{Cu}_{2}\text{O}$ Array before and after thermal oxidation}

The actual dimensions of $Cu$ and $\text{Cu}_{2}\text{O}$ arrays after the deposition and oxidation varied slightly from the nominal dimensions used during the patterning process. When Cu was oxidized to form $\text{Cu}_{2}\text{O}$, due to the changes in the crystal structure and lattice constant and incorporation of the oxygen atoms into the lattice, the volume expanded, thus increasing side length.

\setcounter{figure}{0}
\renewcommand{\figurename}{Fig.}
\renewcommand{\thefigure}{S\arabic{figure}}

\begin{figure}[h]
    \centering
    \includegraphics[scale=0.28]{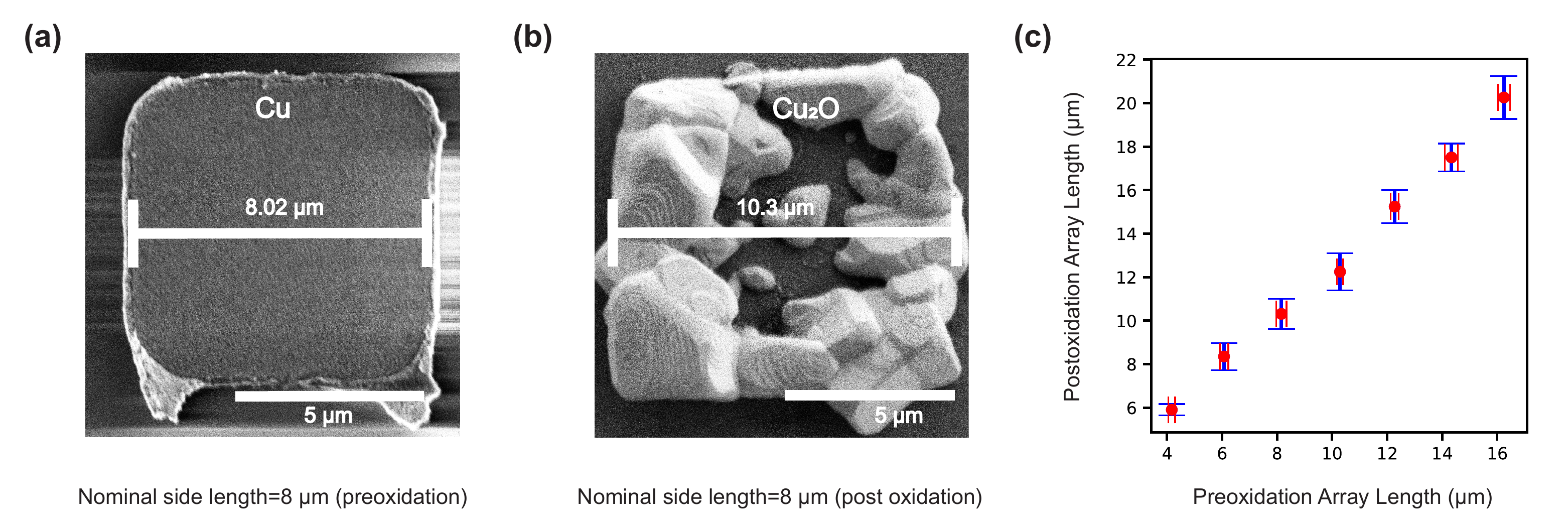}
    \caption{\textbf{Effects of photoresist liftoffs and thermal oxidation on the nominal side length of square copper arrays}. (a) A square copper region having a nominal length (patterning length during lithography) of 8 $\mu$m. After the copper deposition and photoresist liftoff, the actual side length did not change significantly, as evidenced by SEM. (b) After oxidation, the side length of the nominal 8 $\mu$m copper region increased to 10.3 $\mu$m, a 30\% increase in the nominal side length. (c) The lengths of the square copper arrays before and after oxidation. The error bars are the statistical error bars from measurements of different arrays.}
    \label{SideLengthCompare}
\end{figure}

\begin{figure}[!htpb]
    \centering
    \includegraphics[scale=0.26]{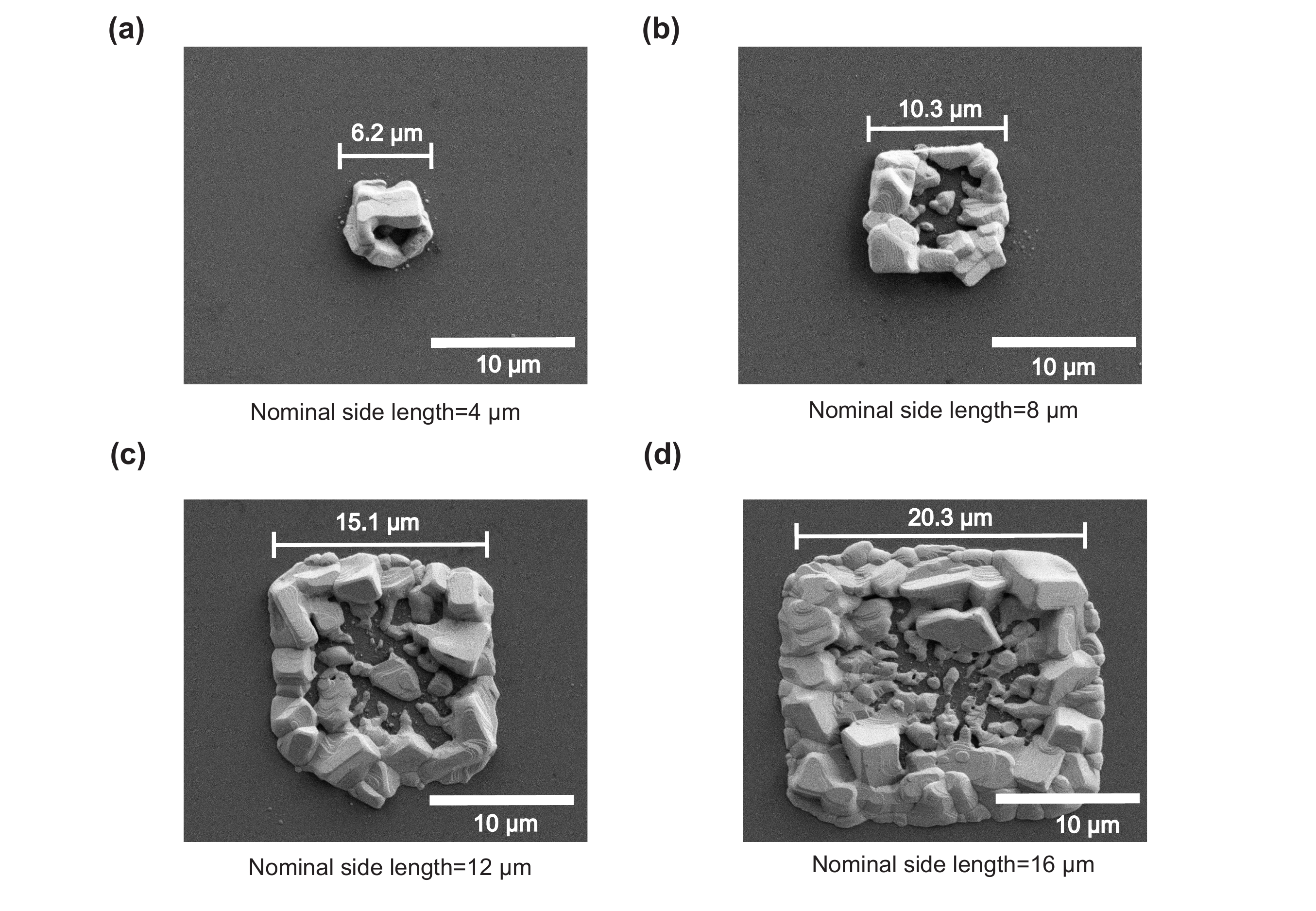}
    \caption{\textbf{Variation of square array side length before and after oxidation}. (a), (b), (c), and (d) show the actual side length of the square $\text{Cu}_{2}\text{O}$ array having nominal side lengths of 4 $\mu$m, 8 $\mu$m, 12 $\mu$m, 16 $\mu$m after oxidation, which showed around 30\% increase from the nominal value. }
    \label{SideLengthforAllArrays}
\end{figure}

\section{Extracting exciton effective temperature from Phonon-assisted transition}

\begin{figure}[h]
    \centering
    \includegraphics[scale=0.4]{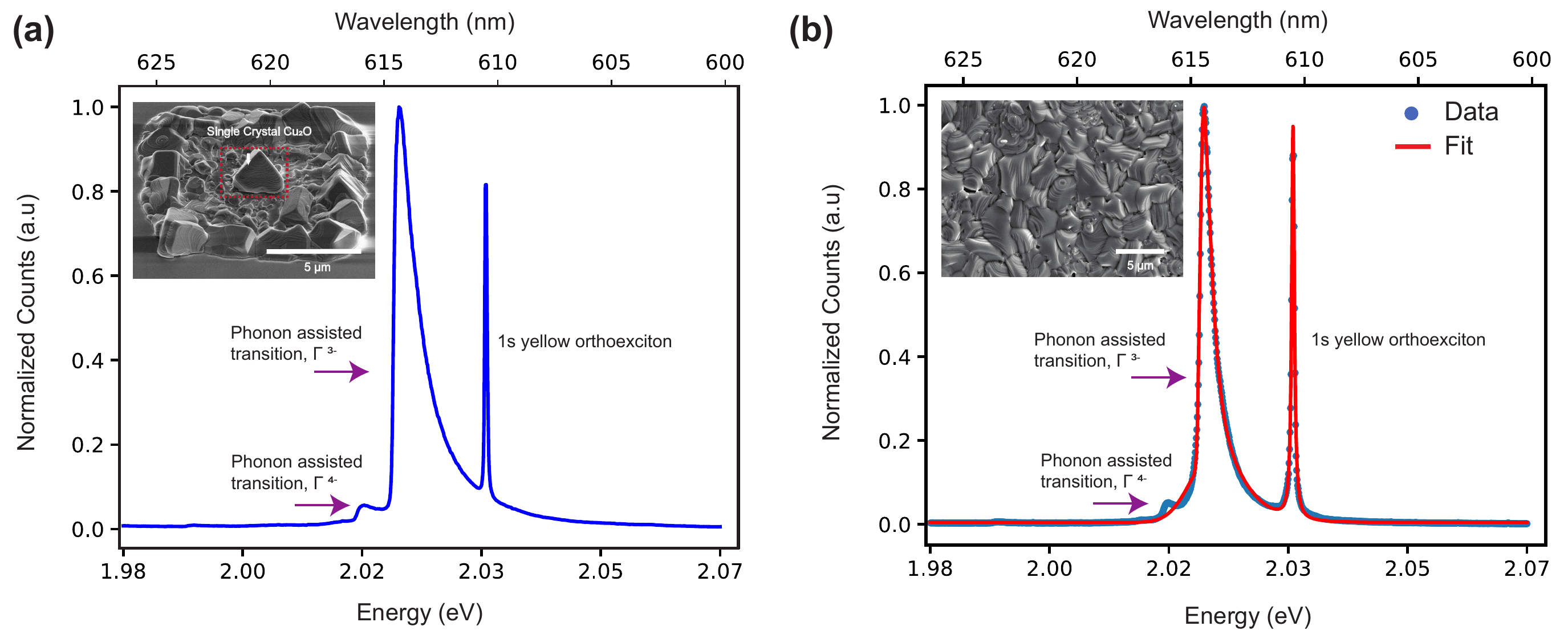}
    \caption{\textbf{Phonon replicas and 1s orthoexciton transition from array and thin-film sample.} (a) Phonon-assisted transitions and 1s orthoexciton from the single-crystal $\text{Cu}_{2}\text{O}$ at the center of the square array (marked with red box). Photoluminescence emission collected from this sample showed a very strong and narrow 1s orthoexciton peak compared to any other position on the array. (b) Phonon-assisted transitions and 1s orthoexciton from the thin-film demonstrated very high count and narrow linewidth in 1s orthoexciton transition. The entire spectrum was fitted using a Maxwell-Boltzmann distribution function convoluted with a Gaussian function to extract the effective temperature of excitons. }
    \label{Sharp1sPeak}
\end{figure}

The $\Gamma_3^-$ phonon-assisted transition has the shape of a Maxwell-Boltzmann distribution function given by the following equation
\begin{equation}
    I(E)\propto\sqrt{E-E_i}\exp{[-(E-E_i)/k_BT_{\rm eff}]}
\end{equation}
To account for the instrument response function of the spectrometer (which has the shape of a Gaussian function), we convolved the Maxwell-Boltzmann distribution function with a Gaussian function and used this to fit the spectrum of phonon-assisted transition and extract exciton effective temperature $T_{eff}$. The convolved function has the following form
\begin{equation}
    I(E)=A(\sqrt{E-E_i}\exp{[-(E-E_i)/k_B T_{\rm eff}]})\ast (\exp{[-(E-E_i)^{2}/2\sigma^{2}])/\sigma\sqrt{2\pi}}
    \label{Eqn2}
\end{equation}
The exciton effective temperature $T_{\rm eff}$ was determined as a fit parameter from Eqn.\ref{Eqn2}. We also used a Lorentzian function to fit the sharp 1s exciton peak. This methodology was used to extract the effective exciton temperature from all phonon-assisted transition peaks. As an example, the raw spectrum, fitted spectrum, and extracted $T_{\rm eff}$ of a thin-film sample (not arrays) are shown in Fig.\ref{Sharp1sPeak} (b). The extracted $T_{\rm eff}$ for thin-film was around 12.07 K.

\begin{figure}[h!]
    \centering
    \includegraphics[width=18cm]{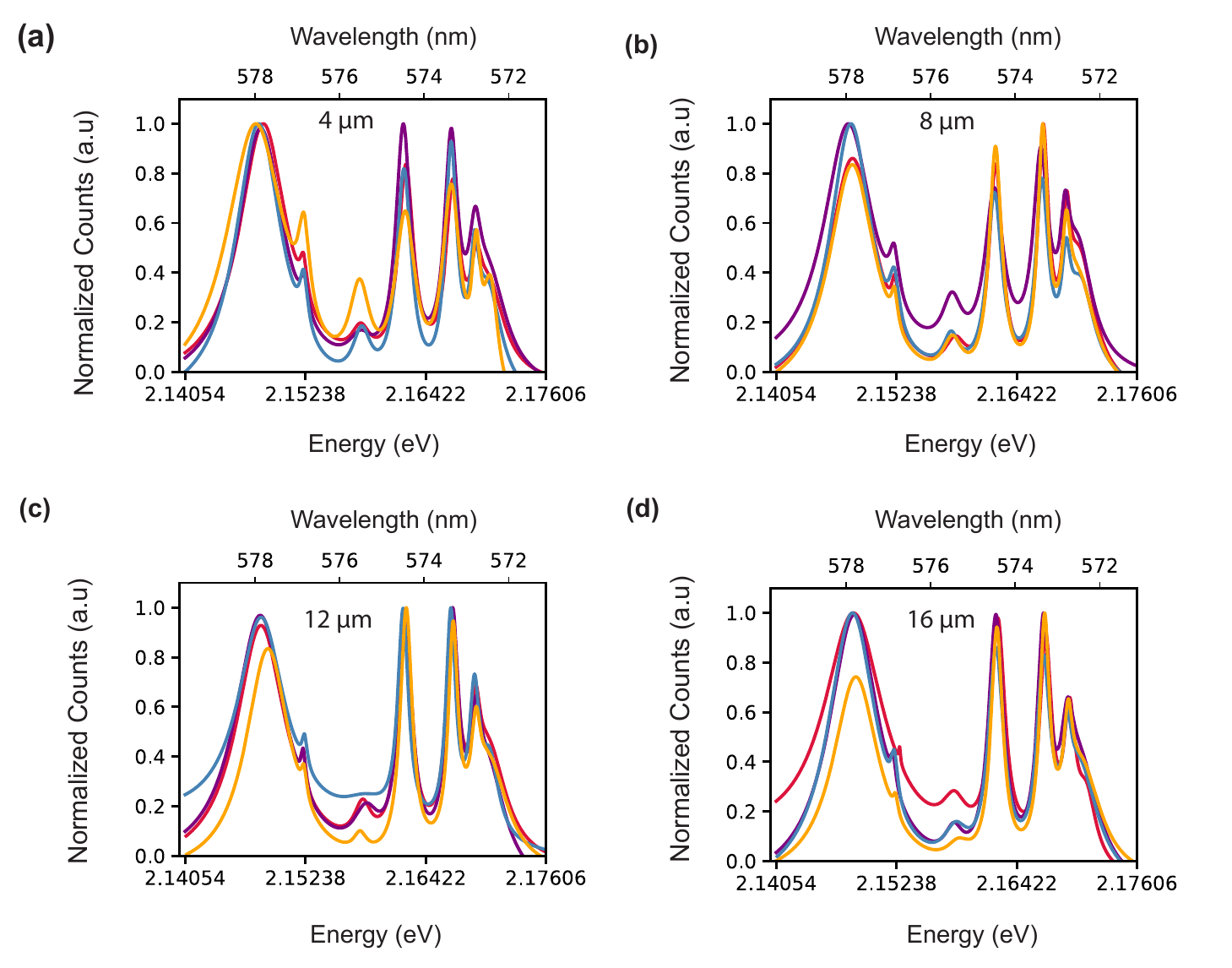}
    \caption{\textbf{Fitted PL spectrum for yellow exciton series for arrays having a nominal side length of 4 $\mu$m, 8 $\mu$m, 12 $\mu$m, and 16 $\mu$m.} For each case, the PL spectrum differed from array to array having the same size. This indicates the non-uniformity of the surface of the sample. Fitting errors are less than 0.1\%.}
    \label{PL_stat_array}
\end{figure}

\begin{figure}[h!]
    \centering
    \includegraphics[width=19cm]{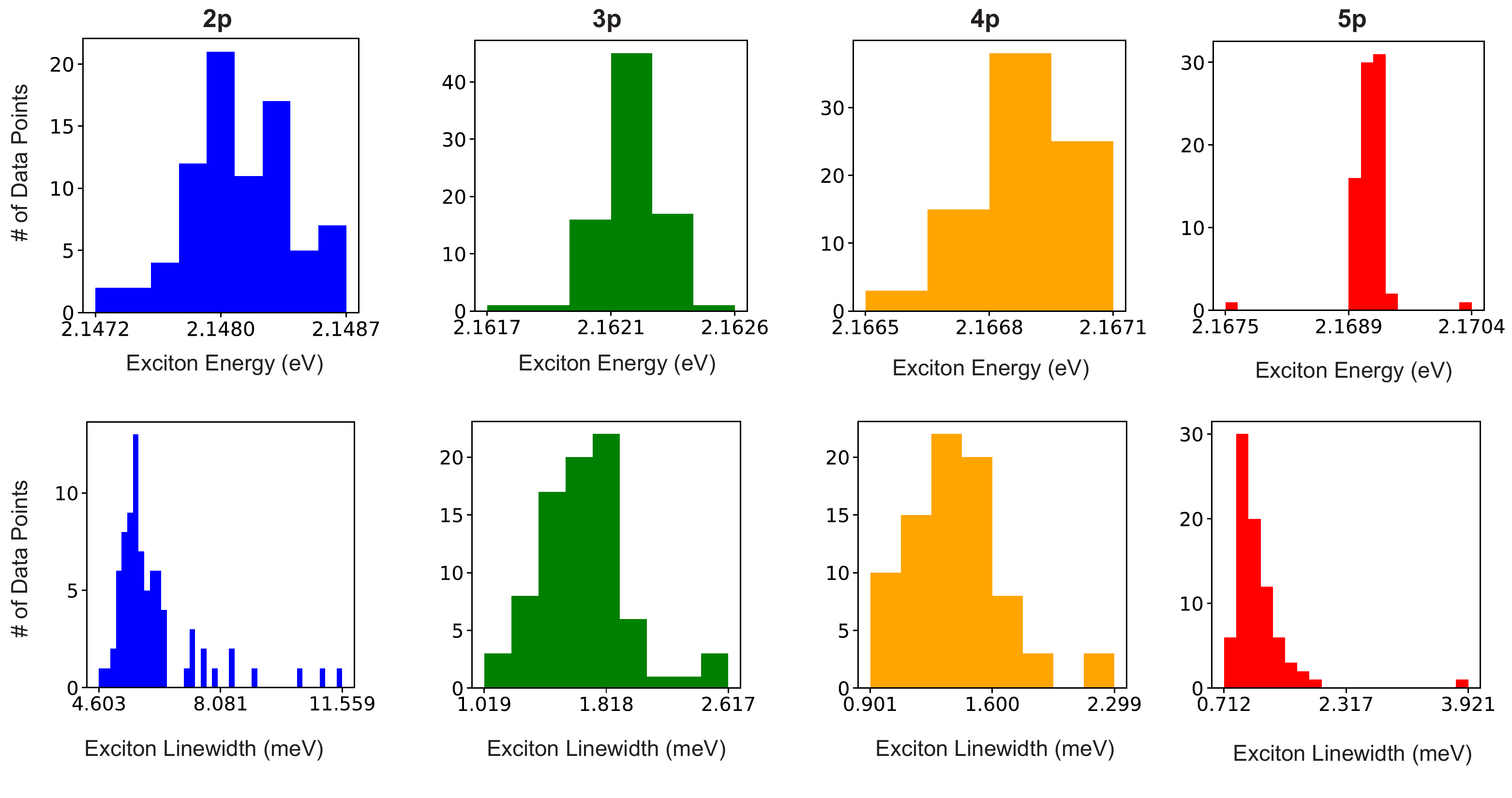}
    \caption{\textbf{Distribution of exciton energies and linewidths for 2p, 3p, 4p and 5p exciton from 80 measurements}}
    \label{histogramofenergies}
\end{figure}

\end{document}